\documentclass[aps,prl,twocolumn,superscriptaddress,nobalancelastpage]{revtex4-2}
\usepackage{iftex}
\usepackage[russian,english]{babel}
\usepackage{preamble}

\begin{document}
\title{Dissipation--Sensitivity Trade-Off in Dissipative Bosonic Systems} 
\author{Kohei Yoshimura}
\email{kohei.yoshimura@riken.jp}
\affiliation{Nonequilibrium Quantum Statistical Mechanics RIKEN Hakubi Research Team, Pioneering Research Institute (PRI), RIKEN, 2-1 Hirosawa, Wako, Saitama 351-0198, Japan}
\affiliation{Universal Biology Institute, The University of Tokyo, 7-3-1 Hongo, Bunkyo-ku, Tokyo 113-0033, Japan}
\author{Ryusuke Hamazaki}
\affiliation{Nonequilibrium Quantum Statistical Mechanics RIKEN Hakubi Research Team, Pioneering Research Institute (PRI), RIKEN, 2-1 Hirosawa, Wako, Saitama 351-0198, Japan}
\affiliation{RIKEN Center for Interdisciplinary Theoretical and Mathematical Sciences (iTHEMS), RIKEN, 2-1 Hirosawa, Wako, Saitama 351-0198, Japan}
\date{\today}

\begin{abstract}
We discover a universal trade-off between the dissipation required to maintain a quantum state more sensitive than a thermal one and the sensitivity gain for a driven single-mode bosonic system under a thermal environment. 
Specifically, for Gaussian states, we rigorously derive a lower bound on the entropy production rate in the steady state in terms of the quantum Fisher information gain relative to a thermal reference state.
Numerical investigations of Kerr nonlinearities further support that these bounds can hold beyond the Gaussian regime. 
Our results establish a new connection between steady-state thermodynamics and quantum metrological performance.
\end{abstract}

\maketitle

\noindent\textit{Introduction}.---
Maintaining high performance realized out of equilibrium inevitably requires energy dissipation~\cite{oono1998steady,seifert2012stochastic,horowitz2015energy,horowitz2017minimum}. 
With the rapid development of nonequilibrium thermodynamics in this century, this fact has been proved in various situations such as autonomous clocks~\cite{erker2017autonomous,schwarzhans2021autonomous}, particle detectors~\cite{schwarzhans2026quantum}, and responses to external signals~\cite{sartori2014thermodynamic,sartori2015free}.
In particular, thermodynamic uncertainty relations~\cite{barato2015thermodynamic,gingrich2016dissipation,horowitz2020thermodynamic,dechant2018multidimensional,liu2020thermodynamic,van2022thermodynamics,yoshimura2021thermodynamic,brandner2025thermodynamic,yoshimura2026quasiprobability} embody the inevitability of dissipation in a highly general way by giving universal trade-off relations between the thermodynamic cost, i.e., entropy production, and precision of currents. 

Quantum metrology raises a related question: how much dissipation is required to keep a sensitive probe ready for use?
A bosonic squeezed state enables quantum estimation with higher precision than the standard quantum limit~\cite{caves1981quantum,schnabel2017squeezed,ganapathy2023broadband} and
has been experimentally realized in various systems~\cite{slusher1985observation,wu1986generation,meekhof1996generation,wollman2015quantum,pirkkalainen2015squeezing,vahlbruch2016detection,kamba2025quantum}; moreover, stationary preparation is studied theoretically~\cite{kronwald2013arbitrarily,kronwald2014dissipative,dicandia2023critical} and experimentally~\cite{wollman2015quantum,lei2016quantum}. 
Nevertheless, the relationship between the benefits of being out of equilibrium, namely, sensitivity improvements brought by external control, such as squeezing, and the thermodynamic cost of maintaining them has not been examined. 
Finding such a trade-off relation between sensitivity improvement and dissipation is not only practically useful for evaluating probe performance in quantum estimation, but can also be conceptually important for highlighting the broader role of entropy production in constraining nonequilibrium performance. 
Note that previous studies have explored thermodynamic aspects of bosonic quantum sensing from various perspectives~\cite{galve2009nonequilibrium,zagoskin2012heat,lipkabartosik2018thermodynamic,chu2022thermodynamic,dooley2022comment,chu2022notes,chu2022erratum,wang2018coherence,wang2022the,chitgopekar2026thermal}; however, none of them answers how much dissipation is needed to maintain a beneficial quantum probe, or how that dissipation relates to sensitivity improvement. 

\begin{figure}[t]
    \centering
    \includegraphics[width=.95\linewidth]{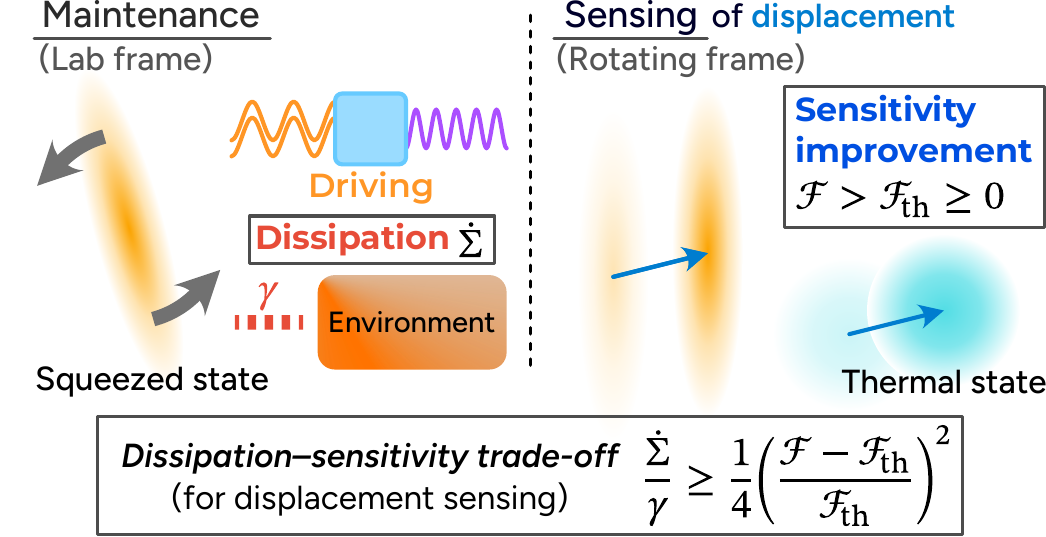}
    \caption{We consider a single-mode bosonic system kept out of equilibrium to possess higher estimation performance. 
    We derive a trade-off between the dissipation required to maintain the sensitive state and the increase in its sensitivity over a thermal state. 
    For displacement sensing, the entropy production rate is bounded quadratically by the quantum Fisher information gain, while linear bounds are obtained for rotation and squeezing sensing. }
    \label{fig:concept}
\end{figure}

In this Letter, we establish a universal dissipation--sensitivity trade-off for a single-mode bosonic probe maintained in a thermal environment (Fig.~\ref{fig:concept}). 
The trade-off relation reads 
\begin{equation}
    \frac{\epr}{\gamma}
    \geq \frac{1}{4}
    \bigg(\frac{\qfi-\qfi_{\therm}}{\qfi_{\mathrm{th}}}\bigg)^2
    \;\;\:\text{or}\quad
    \frac{\epr}{\gamma}
    \geq \frac{1}{4}(\qfi-\qfi_{\therm}),
    \label{eq:intro}
\end{equation}
where the first one applies to displacement sensing and the other to rotation and squeezing sensing. 
We quantify sensitivity using the quantum Fisher information (QFI), denoted by $\qfi$, which gives the ultimate precision in parameter estimation~\cite{giovannetti2011advances,braunstein1994statistical,paris2009quantum}. 
Since a thermal state can have nonzero QFI, its improvement, $\qfi-\qfi_{\therm}$, rather than the QFI itself, plays the central role. 
The inequalities provide a quantitative relationship between this sensitivity improvement and the necessary dissipation, quantified by the entropy production rate (EPR) $\epr$ over the decay rate $\gamma$. 

We derive these inequalities rigorously for Gaussian steady states, a central setting for continuous-variable quantum metrology with experimental accessibility and analytical tractability~\cite{weedbrook2012gaussian,fadel2025quantum}, in the Lindbladian dynamics with a quadratic Hamiltonian and linear jump operators satisfying local detailed balance.
We numerically confirm the tightness of the bounds and also investigate non-Gaussian situations induced by the Kerr nonlinearity.
The numerical results show that the equalities can be approximately obtained in metrologically preferable situations, and indicate that the bounds persist in the non-Gaussian setting, while their general validity remains open to future study. 

\noindent\textit{Dynamics and thermodynamics}.---
We consider a single-mode bosonic system described by the annihilation and creation operators $a$ and $a^\dagger$, which satisfy $[a,a^\dagger]=1$. 
We set $\hbar=1$ and define the quadrature operators in the phase space by $x=(a+a^\dagger)/\sqrt{2}$ and $p=-i(a-a^\dagger)/\sqrt{2}$, so we have $[x,p]=i$. For notational convenience, we introduce $\rr=(x,p)^\top$ and the matrix 
$\Omega = \begin{pmatrix}
0&1\\-1&0
\end{pmatrix}$, 
which simplify the commutation relation as $[\rr,\rr^\top]=i\Omega$. 

The dynamics are assumed to be time-independent in the rotating frame with respect to the reference Hamiltonian $\ham_0=\omega \ada$, modeled by the Lindblad equation~\cite{lindblad1976generators,gorini1976completely,breuer2002theory,milburn1981production,collett1984squeezing} 
\begin{equation}
    \begin{split}
        &\frac{d\rho}{dt}=-i[\ham,\rho] 
        + \diss(\rho),\\
        &\diss(\rho) 
        = \gamma(\bar{n}+1)\diss[a](\rho) 
        + \gamma\bar{n}\diss[a^\dagger](\rho). 
    \end{split} \label{eq:lindblad}
\end{equation}
Here, $\rho=\rho(t)$ is the density operator, $\ham$ is the Hamiltonian, $\gamma>0$ is the decay rate, $\bar{n}=\frac{1}{e^{\beta\omega}-1}$ is the thermal population of the environment at inverse temperature $\beta$, and $\diss[c](\rho)=c\rho c^\dagger -\frac{1}{2}\{c^\dagger c,\rho\}$. 
While Eq.~\eqref{eq:lindblad} is time-independent in the rotating frame, the lab-frame dynamics are described by $d\rho_{\lab}/dt=-i[\ham_0+\ham_{\mathrm{drv}}(t),\rho_{\lab}] + \diss(\rho_\lab)$, involving a time-dependent driving $\ham_{\mathrm{drv}}(t)= U_0(t)\ham U_0(t)^\dagger$, where $ U_0(t)=e^{-i\ham_0t}$~\cite{milburn1981production,collett1984squeezing}. 
The density operators are linked via $\rho_{\lab}(t)= U_0(t)\rho(t) U_0(t)^\dagger$. 
Therefore, the steady state $\rho_\infty$ of Eq.~\eqref{eq:lindblad} is a periodic nonequilibrium steady state in the lab frame, and $\rho_{\lab}(t)$ converges in the long-time limit to $ U_0(t)\rho_\infty U_0(t)^\dagger$. 

We establish the thermodynamic framework of this system by defining the energy as $E(t)=\tr\big(\rho_\lab(t)(\ham_0+\ham_\drv(t))\big)$. 
The work and heat flows are identified as $\dot{W}(t)=\tr\big(\rho_\lab(t)\dot{\ham}_\drv(t)\big)$ and $\dot{Q}(t)=\tr\big(\dot{\rho}_\lab(t)(\ham_0+\ham_\drv(t))\big)$ to fulfill the first law $\frac{d}{dt}E(t) = \dot{W}(t)+\dot{Q}(t)$~\cite{alicki1979quantum,spohn1978entropy}. 
This identification allows us to obtain a well-defined expression for the entropy production rate (EPR) at the steady state. 
Since the von Neumann entropy becomes invariant in the steady state because $-\tr\big(\rho_\lab(t)\ln\rho_\lab(t)\big)=-\tr(\rho_\infty\ln\rho_\infty)$ due to unitary invariance, we can define the steady-state EPR by $\epr = -\beta\dot{Q}$. 
Using the first law and the Lindblad equation, we finally obtain the expression
\begin{align}
    \epr 
    = \gamma\beta\omega\big(\langle\ada\rangle-\bar{n}\big), 
    \label{eq:epr}
\end{align}
which is proved in the End Matter (EM). 
Note that we are using the most standard definition of entropy production~\cite{landi2021irreversible} rather than the Wigner-function-based one~\cite{santos2017wigner}. 

In the following, we further assume that the Hamiltonian is quadratic as 
\begin{align}
    \ham = \frac{1}{2}\rr^\top \HH\rr + \hh^\top\rr, 
\end{align}
where $\HH$ is a symmetric $2\times 2$ real matrix and $\hh$ is a two-dimensional real vector (they can be zero). 
Under this setting of a quadratic Hamiltonian and linear dissipators in Eq.~\eqref{eq:lindblad}, the stable steady state becomes Gaussian~\cite{mancini2007optimal,rudnicki2018stabilizable}, which is designated by the mean vector and covariance matrix, 
\begin{align}
    \rbar = \langle \rr\rangle,\quad 
    \VV = \frac{1}{2}\big\langle \{\rr,\rr^\top\}\big\rangle-\rbar \,\rbar^\top.
\end{align}
Here, $\langle\cdot\rangle$ indicates an expectation value and $[\{\rr,\rr^\top\}]_{ij}=\{r_i,r_j\}$~\cite{weedbrook2012gaussian}. 
Some other basic features of a Gaussian state (and its definition) are provided in the EM for self-containedness. 

Note that we intentionally avoid using squeezed baths~\cite{breuer2002theory,manzano2016entropy}. 
A squeezed bath can realize a highly sensitive squeezed state for free (i.e., without dissipation) if we define entropy production based on the corresponding Lindblad equation; however, producing a squeezed bath requires thermodynamic cost that cannot be accounted for by the system information~\cite{singh2020performance}. 
Thus, we focus on the thermal bath in order to investigate the trade-off between dissipation and sensitivity in a simple and complete way. 

\noindent\textit{Quantum Fisher Information}.---
Next, we introduce the quantum Fisher information (QFI) for an arbitrary state $\rho$. 
When restricted to unitary operations such as $\rho\mapsto \rho_\theta=e^{-i\theta X}\rho e^{i\theta X}$, the QFI is defined for $\rho$ and the Hermitian operator $X$ as $\qfi_\rho(X)=\tr(\rho L^2)$, where $L$ is an operator that satisfies $-i[X,\rho] = \frac{1}{2}\{\rho,L\}$~\cite{braunstein1994statistical,paris2009quantum} (see also the supplemental material (SM)~\cite{smArxiv}). 
This quantity sets an attainable lower bound on the variance of any unbiased estimator $\theta_{\mathrm{est}}$ of the unknown parameter $\theta$ encoded in the operation. 
Such a bound, known as the quantum Cram\'{e}r--Rao bound and given as $\Var(\theta_{\mathrm{est}})\geq 1/\qfi_\rho(X)$, accounts for the role of the QFI as the sensitivity in state $\rho$ when it is used as a probe. 

\begin{figure}
    \centering
    \includegraphics[width=.8\linewidth]{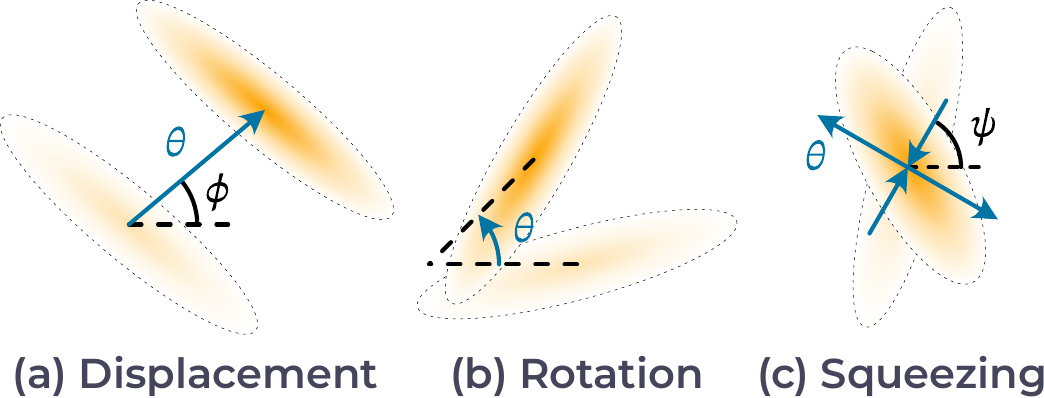}
    \caption{We focus on three sensing protocols in the phase space: (a) displacement in angle $\phi$, (b) rotation about the origin, and (c) squeezing in angle $\psi$. Note that while $\theta$ is the estimated parameter, $\phi$ and $\psi$ are fixed parameters that designate the operation. }
    \label{fig:operations}
\end{figure}

To be concrete, in this study, we focus on the three most representative operations in the phase space~\cite{fadel2025quantum}: (a) displacement in the angle $\phi$, (b) rotation about the origin, and (c) squeezing in the angle $\psi$ (see Fig.~\ref{fig:operations}). 
In a single-mode system, these operations are induced by Hermitian operators $X_\dis^\phi=\nn_\phi^\top\Omega\rr$, $X_\rot=\frac{1}{2}\rr^\top\rr$, and $X_\sq^\psi = \frac{1}{2}\rr^\top \QQ_\psi\rr$, where $\nn_\phi=(\cos\phi,\sin\phi)^\top$ and 
$\QQ_\psi = \begin{pmatrix}
\sin(2\psi) & -\cos(2\psi) \\
-\cos(2\psi) & -\sin(2\psi)\end{pmatrix}$~\cite{smArxiv}. 

If $\rho$ is Gaussian, the QFI is known to have a closed-form expression~\cite{pinel2013quantum,monras2013phase,jiang2014quantum}. In particular, for the three operations, we have $\qfi_\rho(X_\dis^\phi) = \nn_{\phi}^\top \VV^{-1}\nn_\phi$, $\qfi_\rho(X_\rot) = \frac{\rbar^\top\VV\rbar}{\det(\VV)} + \frac{[\tr(\VV)]^2-4\det(\VV)}{\det(\VV)+1/4}$, and $\qfi_\rho(X_\sq^\psi) = \frac{\rbar^\top\QQ_\psi\VV\QQ_\psi\rbar}{\det(\VV)} + \frac{[\tr(\VV\QQ_\psi)]^2+4\det(\VV)}{\det(\VV)+1/4}$, 
where the mean and covariance are evaluated at $\rho$~\cite{smArxiv}. 
For simplicity, we will write the QFIs at $\rho=\rho_\infty$ as $\qfi_\dis(\phi)$, $\qfi_\rot$, and $\qfi_\sq(\psi)$. 

When discussing a trade-off between the steady-state EPR and the QFI in the steady state, we must be aware that the QFI can be nonzero even if the state is an equilibrium state, where the EPR vanishes. 
This suggests that we need to focus on the improvement of the QFI relative to a thermal state rather than the QFI itself. 
Such a thermal reference state is always uniquely determined for any Gaussian state~\cite{weedbrook2012gaussian,serafini2023quantum} (see also the EM). 
In this two-dimensional case, the thermal state $\rho_{\therm}$ for an arbitrary Gaussian state $\rho$ with covariance $\VV$ has zero mean $\rbar_{\therm}=0$ and covariance matrix $\VV_{\therm}=\sqrt{\det(\VV)}\id$, where $\id$ is the identity matrix. 
Thus, the thermal offsets are given as $\qfi_{\rho_{\therm}}(X_\dis^\phi) = \frac{1}{\sqrt{\det(\VV)}}$, $\qfi_{\rho_{\therm}}(X_\rot) = 0$, and $\qfi_{\rho_{\therm}}(X_\sq^\psi) = \frac{4\det(\VV)}{\det(\VV)+1/4}$. 
We indicate the QFIs at the thermal state corresponding to $\rho_\infty$ by a superscript $\therm$ and omit inessential arguments; e.g., the displacement QFI is written as $\qfi_\dis^\therm$. 

\noindent\textit{Dissipation-sensitivity trade-off}.---
Trade-off relations are derived from the expression 
\begin{align}
    \frac{\epr}{\gamma} = \bigg(\lVert\rbar\rVert^2 
    + \frac{[\tr(\VV)]^2-4\det(\VV)}{\tr(\VV)}\bigg)
    \arcoth\bigg(\frac{4\det(\VV)}{\tr(\VV)}\bigg). 
    \label{eq:epr_formula}
\end{align}
To our knowledge, this formula is unknown in the literature. 
Interestingly, it connects ``results,'' i.e., the entropy production per decay time ($1/\gamma$), the mean, and the covariance, while excluding ``causes'' such as $\beta$, $\omega$, $\HH$, or $\hh$.
We will prove the equality in a generalized ($N$-mode) form in the EM, where we also verify the non-negativity and that the argument of $\arcoth$ is not smaller than one. 

We mention that this expression motivates us to split the EPR into two parts: one due to the displaced center $\rbar$ and the other stemming from the squeezed distribution. 
Specifically, we will use the latter one, explicitly given as $\epr_{\sq}=\gamma\frac{[\tr(\VV)]^2-4\det(\VV)}{\tr(\VV)}
\arcoth\big(\frac{4\det(\VV)}{\tr(\VV)}\big)$. 
This partial EPR is never greater than the total EPR and vanishes when $\VV$ is proportional to the identity, i.e., the steady state is isotropic. 

Based on Eq.~\eqref{eq:epr_formula}, we obtain our main result, which we name the \textit{dissipation--sensitivity trade-off}: 
\begin{subequations}\label{eq:inequalities}
    \begin{align}
        \frac{\epr_\sq}{\gamma}
        &\geq \frac{1}{4}
        \bigg(
            \frac{\qfi_\dis(\phi)-\qfi_{\dis}^{\therm}}
                {\qfi_{\dis}^{\therm}}
        \bigg)^2, \label{eq:ineq_dis}\\
        \frac{\epr}{\gamma}
        &\geq \frac{\qfi_\rot}{4},\\
        \frac{\epr}{\gamma}
        &\geq \frac{\qfi_\sq(\psi)-\qfi_\sq^{\therm}}{4}. 
    \end{align}
\end{subequations}
These inequalities lead to a simple physical conclusion that maintaining higher sensitivity requires greater thermal dissipation. 
They hold for any $\phi$ and $\psi$, even though $\qfi_{\dis}(\phi)-\qfi_\dis^\therm$ can be negative~\footnote{With $v_+\geq v_-$ being the eigenvalues of $\VV$, the QFI falls in the interval $[1/v_+,1/v_-]$, while $\qfi_\dis^{\therm}=1/{\sqrt{v_+v_-}}$. Thus, we get $\min_\phi\qfi_\dis(\phi)\leq\qfi_\dis^{\therm}\leq \max_\phi\qfi_\dis(\phi)$. Due to the inequality between the arithmetic mean and the geometric mean, we also have $[\max_\phi\qfi_\dis(\phi)+\min_\phi\qfi_\dis(\phi)]/2\geq \qfi_\dis^{\therm}$, which means the lower bound in Eq.~\eqref{eq:ineq_dis} is maximized when $\phi=\arg\max_{\phi'}\qfi_\dis(\phi')$ (seen by drawing a parabola). }. 

The physical meaning of the inequalities is further clarified when we connect them to possible estimation error. 
In terms of an estimator $\theta_{\mathrm{est}}$ of the parameter $\theta$, for instance, Eq.~\eqref{eq:ineq_dis} turns into 
\begin{align}
    \frac{\epr_\sq}{\gamma}
    &\geq \frac{1}{4}
    \bigg(
        \frac{\Var(\theta^*)}{\Var(\theta_{\mathrm{est}})}-1
    \bigg)^2,
\end{align}
if the estimator satisfies $\Var(\theta_{\mathrm{est}})\leq \Var(\theta^*)$, where $\theta^*$ is the optimal estimator when we use the thermal reference state as the probe. 
This inequality is proved by combining Eq.~\eqref{eq:ineq_dis} and the Cram\'{e}r--Rao inequality and using the fact that equality in the Cram\'{e}r--Rao inequality is always attainable. 
Thus, if we try to realize a sensor more sensitive than the thermal state, there is inevitable dissipation depending on the sensitivity improvement. 
The complete proof and similar inequalities for the other two operations are shown in the SM~\cite{smArxiv}. 

For displacement sensing, Eq.~\eqref{eq:ineq_dis} tells us that dissipation is also imposed by nonclassicality. 
A Gaussian state has a non-positive (thus, nonclassical) $P$-function if and only if the QFI $\qfi_\rho(X_{\mathrm{dis}}^\phi)$ is greater than $2$ for some $\phi$~\cite{kwon2019nonclassicality,serafini2023quantum}. Z
On the other hand, uncertainty principle asserts $\det(\VV)\geq 1/4$, which means $\qfi_{\dis}^{\therm}\leq 2$. 
Thus, the lower bound in Eq.~\eqref{eq:ineq_dis} can always be nonzero if $\rho_\infty$ is nonclassical, which means nonclassicality requires dissipation. 
Note that the inverse is not always true because we may have $2 \geq\qfi_{\dis}(\phi)\geq \qfi_{\dis}^{\therm}$ for all $\phi$. 

We explain the central idea of the proof of the inequalities while skipping the detailed proof to the SM~\cite{smArxiv}. 
The key inequality is $\arcoth(x)\geq 1/x$ valid for $x\geq 1$. 
Applying this inequality to Eq.~\eqref{eq:epr_formula} gives 
\begin{align}
    \frac{\epr}{\gamma}
    \geq \frac{\tr(\VV)}{4\det(\VV)}\lVert\rbar\rVert^2
    + \frac{[\tr(\VV)]^2-4\det(\VV)}{4\det(\VV)}. \label{eq:epr_lowerbound}
\end{align}
By comparing the QFI improvement with the right-hand side, we will find the desired inequalities. 
For example, the rotation QFI is bounded as 
\begin{align}
    \qfi_\rot
    \leq \frac{\tr(\VV)}{\det(\VV)}\lVert\rbar\rVert^2
        + \frac{[\tr(\VV)]^2-4\det(\VV)}{\det(\VV)},
\end{align}
where the upper bound is exactly four times the lower bound in Eq.~\eqref{eq:epr_lowerbound} (this inequality is proved using $\VV\leq \tr(\VV)\id$). 

\begin{figure}[t]
    \centering
    \includegraphics[width=\linewidth]{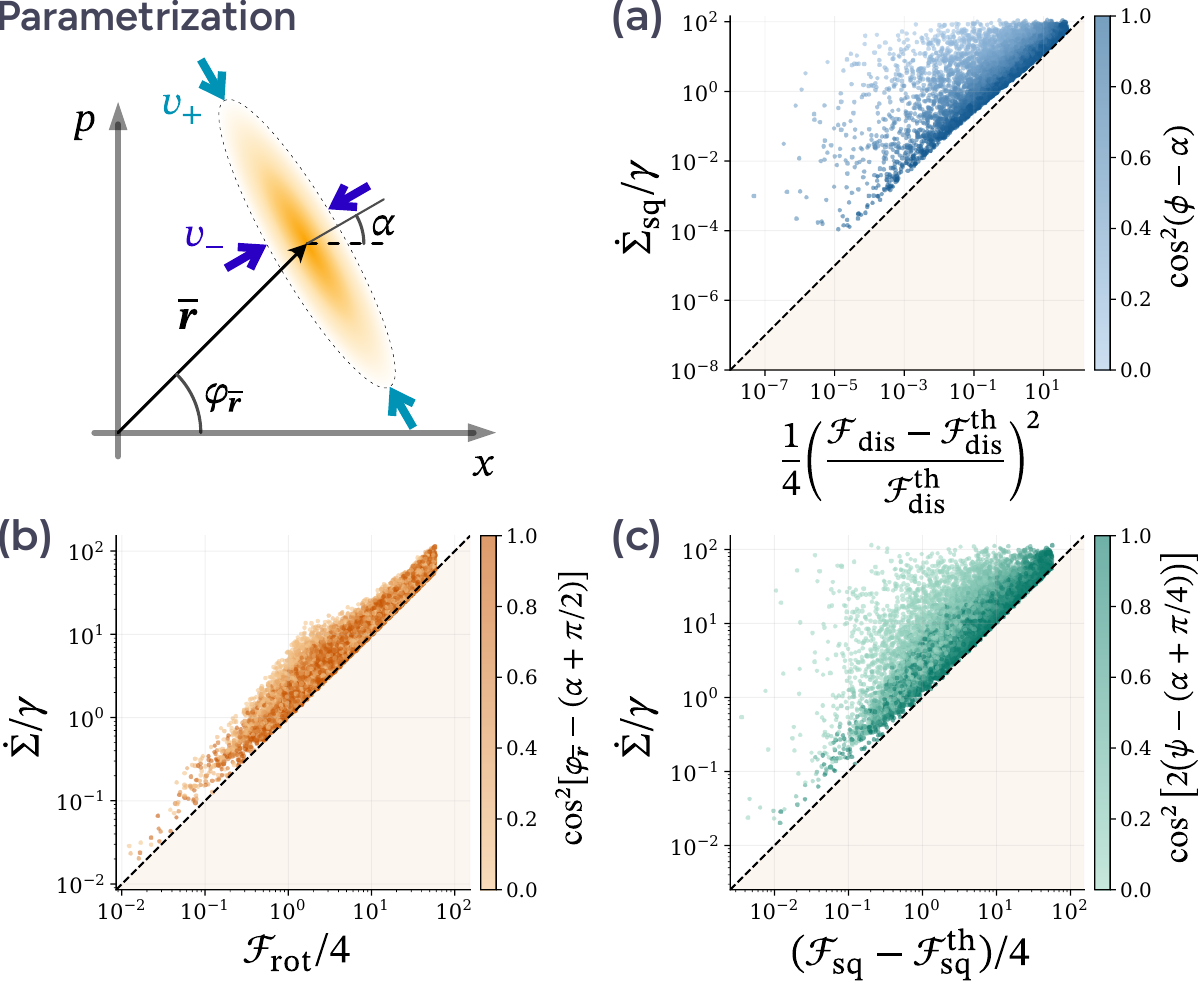}
    \caption{Numerical verification of inequalities~\eqref{eq:inequalities}. 
    In the top left, we show the parametrization of steady states. 
    We plot the EPR and the QFI bounds at randomly generated 8000 steady states for (a) displacement, (b) rotation, and (c) squeezing. 
    No violation was observed. 
    The dots are colored based on the similarity between (a) squeezed angle $\alpha$ and displacement angle $\phi$, (b) anti-squeezed angle $\alpha+\pi/2$ and $\rbar$'s direction, and (c) a middle angle $\alpha+\pi/4$ and squeezing angle $\psi$ (up to a $\pi/2$ difference). 
    When the two angles are close, the inequality becomes tight. 
    For more details, including the choice of sampling parameters, see the SM~\cite{smArxiv}. }
    \label{fig:bounds_gauss}
\end{figure}

\noindent\textit{Numerical validation}.---
We verify our dissipation--sensitivity trade-off by randomly generating steady states designated by $\rbar$ and $\VV$ (for the validity of the following parametrization, see the EM). 
While $\rbar$ can be an arbitrary two-dimensional real vector, $\VV$ has to satisfy some conditions. 
As proved in the EM, any steady-state matrix $\VV$ is expressed as follows: 
\begin{equation}\label{eq:parametrization}
    \begin{split}
        &\VV 
        = \RR(\alpha)\begin{pmatrix}
            v_- & 0 \\ 0 & v_+
        \end{pmatrix}\RR(\alpha)^\top,\\
        &v_\pm = \bigg(\bar{n}+\frac{1}{2}\bigg)\cosh(2s)e^{\pm 2s}\quad(s\geq0). 
    \end{split}
\end{equation}
where $\RR(\alpha)$ is the rotation matrix by angle $\alpha$. 
The angle $\alpha$ indicates the squeezed direction. 
The thermal population $\bar{n}$ appears in $v_\pm$ reflecting the fact that the associated Gaussian state is a steady state of the Lindblad equation. 
Finally, we find that we have five free parameters: $\lVert\rbar\rVert$, angle $\varphi_{\rbar}=\arctan(\bar{p}/\bar{x})$, $\bar{n}$, $\alpha$, and $s$ (see also the Parametrization in Fig.~\ref{fig:bounds_gauss}). 

Figure~\ref{fig:bounds_gauss}(a)--(c) numerically confirms that the 
inequalities in Eq.~\eqref{eq:inequalities} work well. 
The inequalities become tight when the squeezed angle $\alpha$ points in a desirable direction:
(a) for displacement sensing, it should be parallel to the displacement angle $\phi$;
(b) for rotation, if the anti-squeezed angle $\alpha+\pi/2$ is parallel to $\rbar$'s direction $\varphi_{\rbar}$, the bound becomes relatively tighter; and 
(c) squeezing sensing becomes efficient when the squeezing angle $\psi$ is located at the midpoint of the squeezed and anti-squeezed angles, $\alpha+\pi/4$. 
Indeed, these conditions can be intuitively expected to make the sensing effective; thus, this result suggests that metrologically effective setups can also be thermodynamically efficient. 

\begin{figure}[t]
    \centering
    \includegraphics[width=\linewidth]{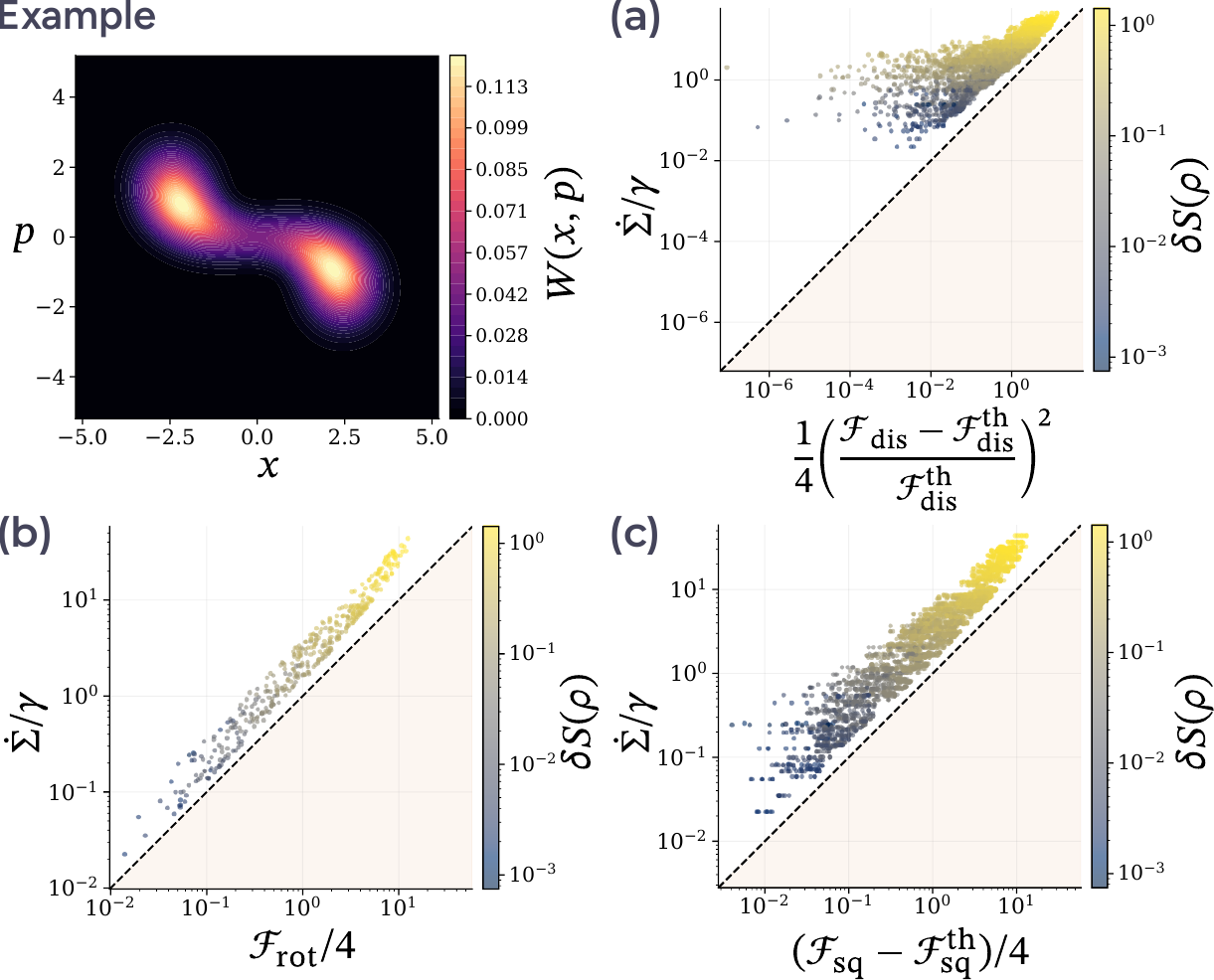}
    \caption{Numerical verification of the dissipation--sensitivity trade-offs for non-Gaussian states (details found in~\cite{smArxiv}). 
    In the top left, the Wigner function of a typical steady state with two peaks is shown.
    Panels (a), (b), and (c) correspond to displacement, rotation, and squeezing bounds. 
    No violation was observed. 
    The dots are colored by the non-Gaussianity measured by the von Neumann entropy decrease from the corresponding Gaussian state with the same mean and covariance as the steady state.     Unlike the Gaussian simulation, we randomly sample Hamiltonian parameters $G$ and $K$ (in addition to $\bar{n}$) rather than $\rbar$ and $\VV$.
    In (a) and (c), we randomly choose eight angles for both $\phi$ and $\psi$, so there are at most eight times as many points as in (b). 
    }
    \label{fig:nongauss}
\end{figure}

\noindent\textit{A non-Gaussian case}.---
So far, we have discussed Gaussian systems, where the Hamiltonian is quadratic, and jump operators are linear. 
One way to go beyond Gaussianity is to add a higher-order term to the Hamiltonian~\cite{lloyd1999quantum,walschaers2021non}. 
To observe non-Gaussianity effects on the inequalities, we adopt the Kerr nonlinearity term $K\adsq a^2$~\cite{drummond1980quantum,kirchmair2013observation}. 
For the Hamiltonian $\ham = \frac{G^*}{2}a^2+\frac{G}{2}\adsq+K\adsq a^2$, we can observe a bimodal steady state when $|G|>\gamma/2$ (see the Example in Fig.~\ref{fig:nongauss}). 

In Fig.~\ref{fig:nongauss}(a)--(c), we plot EPR vs QFI at steady states obtained by numerically solving $-i[\ham,\rho_\infty]+\diss(\rho_\infty)=0$ with $\ham = \frac{G^*}{2}a^2+\frac{G}{2}\adsq+K\adsq a^2$. 
In each case, the inequality is robust to the Kerr nonlinearity. 
We also color the points by their non-Gaussianity measured by the von Neumann entropy decrease $\delta S$ (note that a Gaussian state is the state with the largest von Neumann entropy among all states with the same covariance)~\cite{walschaers2021non}. 
The result suggests that in this case, non-Gaussianity increases not only the QFI but also dissipation, and as a result, the trade-off is not violated. 
More elaborate use of non-Gaussianity could make EPR smaller than the QFI bound, but a decisive conclusion remains open to future research. 

\noindent\textit{Conclusion and outlook}.---
In this work, we discovered the dissipation--sensitivity trade-off~\eqref{eq:inequalities} for a single-mode bosonic system described by a Lindblad equation that is time-independent in a rotating frame. 
It was accomplished by finding a formula~\eqref{eq:epr_formula} that computes the dissipation required to maintain the steady state and connecting it to the state's sensitivity for three different tasks. 
The dissipation--sensitivity trade-offs were numerically validated even in a representative non-Gaussian setup under Kerr nonlinearity. 
It is remarkable that the trade-off relation involves solely well-established quantities, namely the entropy production rate and the quantum Fisher information, and despite its simple form, the bounds can become tight in a wide parameter regime. 
We stress that our results are distinct from well-studied trade-offs such as thermodynamic uncertainty relations and thermodynamic speed limits~\cite{shiraishi2018speed,funo2019speed,ito2020stochastic}. 
Indeed, despite their apparent similarity, the proof of our inequalities is completely different from that of such relations. 

One compelling future direction is to study the scaling of the inequalities. 
In thermodynamic trade-off relations such as thermodynamic uncertainty relations and thermodynamic/quantum speed limits, scaling with system size can be essential~\cite{hamazaki2022speed}. 
Some previous work has shown that quantum systems can use quantum resources, such as coherence, to increase efficiency drastically compared to classical systems~\cite{tajima2021superconducting,funo2025symmetry,kamimura2022quantum,kamimura2023universal,yoshimura2026quasiprobability}. 
The situation is similar in quantum metrology, where the QFI can be increased by entangling probes in a multimode system, leading to high sensitivity unattainable in classical systems~\cite{giovannetti2011advances}. 
Going beyond the current work, which focuses on single-mode systems, multimode extensions of the dissipation--sensitivity trade-off may connect these scaling results. 

\begin{acknowledgments}
    \noindent\textit{Acknowledgments.}---
    K.Y.\ thanks Hiroyasu Tajima, Shunsuke Kamimura, and Artemy Kolchinsky for valuable discussions. 
    The authors are also grateful to Kazuki Yokomizo and Naruo Ohga for fruitful discussions. 
    K.Y.\ used ChatGPT 6 Astra and 5.6 Sol to improve the readability and to write the numerical codes, which he verified himself. He also used QuTiP~\cite{johansson2012qutip} in the implementation. 
    K.Y.\ is supported by the Special Postdoctoral Researchers Program at RIKEN and JSPS KAKENHI Grant Numbers JP25K24775 and JP26K24529. 
    R.H.\ is supported by JSPS KAKENHI Grant Number JP24K16982.
    K.Y.\ and R.H.\ are supported by JST ERATO Grant Number JPMJER2302, Japan.

    \noindent\textit{Data availability.}---
    Numerical codes and CSV data used to generate Figs.~\ref{fig:bounds_gauss} and \ref{fig:nongauss} are available in Ref.~\cite{data}. 
\end{acknowledgments}

%

\clearpage
\section{End Matter}
\subsection{Proof of Eq.~\eqref{eq:epr}}
Because $\rho_{\lab}(t)\to U_0(t)\rho_\infty U_0(t)^\dagger$ for large $t$ and $\ham_{\drv}(t)= U_0(t)\ham U_0(t)^\dagger$, the energy becomes time-independent in the steady state as $E(t)\to \tr\big(\rho_\infty(\ham_0+\ham)\big)$. 
Thus, the first law shows $\epr=-\beta\dot{Q}=\beta\dot{W}=\beta\tr(\rho_\lab(t)\dot{\ham}_{\drv}(t))$. 
Plugging in $\rho_{\lab}=U_0(t)\rho_\infty U_0(t)^\dagger$ and $\ham_{\drv}(t)=U_0(t)\ham U_0(t)^\dagger$, we obtain  
\begin{align}
    \epr=-i\beta\tr\big(\rho_\infty[\ham_0,\ham]\big)
    = -i\beta\tr\big([\ham,\rho_\infty]\ham_0\big), 
    \label{eq:epr_ham}
\end{align}
where we used the cyclic property of the trace in the second equality. 
Since $\rho_\infty$ is the steady-state solution to Eq.~\eqref{eq:lindblad}, we have $i[\ham,\rho_\infty]=\diss(\rho_\infty)$ and thus
\begin{align}
    \epr=-\beta\tr\big(\rho_\infty\diss^\dagger(\ham_0)\big), 
    \label{eq:epr_diss}
\end{align}
where $\diss^\dagger$ is the adjoint of $\diss$. 
Finally, standard calculation of commutation relations yields Eq.~\eqref{eq:epr}. 
Note that, except for the dynamical equation~\eqref{eq:lindblad} and the form of the dissipator, we did not use any assumptions such as Gaussianity of $\ham$ or $\rho$ or the single-mode properties; thus, Eqs.~\eqref{eq:epr_ham} and \eqref{eq:epr_diss} are rather generic expressions of the EPR. 

\subsection{Gaussian states and thermal states}
Here, we consider an $N$-mode bosonic system, which is described by quadrature operators $\rr_N=(x_1,p_1,\dots,x_N,p_N)^\top$. 
Detailed proof of the facts presented here can be found in Ref.~\cite{serafini2023quantum}. 
The canonical commutation relation keeps its simple form $[\rr_N,\rr_N^\top]=i\Omega_N$ with $\Omega_N=\bigoplus_{j=1}^N\Omega$. 
Hereafter, we omit the subscript $N$ for simplicity. 
The mean vector and covariance matrix are naturally defined for $N$-mode systems. 

A density operator $\rho$ is called a Gaussian state if it admits the form 
\begin{align}
    \rho = \frac{e^{-\frac{1}{2}(\rr-\rbar)^\top \KK(\rr-\rbar)}}{\tr[e^{-\frac{1}{2}(\rr-\rbar)^\top \KK(\rr-\rbar)}]}
\end{align}
with $\rbar\in\mathbb{R}^{2N}$ and a symmetric positive-definite $2N\times 2N$ matrix $\KK$. 
The denominator is given by the covariance matrix as $\tr[e^{-\frac{1}{2}(\rr-\rbar)^\top \KK(\rr-\rbar)}]=\sqrt{\det(\VV+\frac{i}{2}\Omega)}$, where the equality $\VV = \frac{i}{2}\coth(\frac{i\Omega \KK}{2})\Omega$ holds. 

When $\rbar=0$ and $\KK$ has the form $\bigoplus_{j=1}^N \nu_j\id_2$ ($\nu_j>0$), we call the state a thermal state. 
If we identify the value $\nu_j$ as inverse temperature times frequency, $\nu_j = \beta'\omega_j$, the state is seen as the Gibbs state $e^{-\beta' \ham_0}/\tr[e^{-\beta'\ham_0}]$ with $\ham_0 = \sum_j\omega_j a_j^\dagger a_j$. 

The Williamson theorem states that every symmetric positive-definite $2N\times 2N$ matrix $\KK$ is decomposed as $\KK=\Smat\KK_d\Smat^\top$ with $\KK_d=\bigoplus_{j=1}^N \nu_j\id_2$, where $\nu_j$ are positive numbers unique to $\KK$ (known as the symplectic eigenvalues) and $\Smat$ is a symplectic matrix, i.e., satisfies $\Smat\Omega\Smat^\top = \Omega$. 
It can further be shown that every Gaussian state $\rho$ is realized by applying a unitary operator(s) to the thermal state corresponding to the symplectic eigenvalues, $\rho_{\therm}=e^{-\frac{1}{2}\rr^\top\KK_d\rr}/\tr\big[e^{-\frac{1}{2}\rr^\top\KK_d\rr}\big]$. 
Because choosing $\rr'=\Smat^\top\rr$ allows $(\rr-\rbar)^\top\KK(\rr-\rbar)=(\rr'-\rbar')^\top \KK_d(\rr'-\rbar')$, the covariance matrix $\VV_{\therm}$ of the thermal state connects to $\VV$ via $\VV=(\Smat^{-1})^\top\VV_{\therm}\Smat^{-1}$. 
Since a symplectic matrix has determinant $1$, we have $\det(\VV_{\therm})=\det(\VV)$. 

In general, the thermal reference state $\rho_\therm$ of a steady state of the driven Lindblad equation differs from the equilibrium state corresponding to the undriven situation, i.e., $\rho_{\mathrm{eq}}=e^{-\beta\ham_0}/\tr[e^{-\beta\ham_0}]$. 
Still, for the single-mode system considered in the main text, it is a natural choice of reference from the thermodynamic perspective because, in the context of work extraction, the reference state is called the passive state and characterizes the maximum work we can extract from the state $\rho$ by a unitary operation~\cite{medina2025anomalous}. 

\subsection{Gaussian dynamics}
The Lindblad equation~\eqref{eq:lindblad} is also extended to $N$-mode cases. 
We generally assume the quadratic Hamiltonian $\ham=\frac{1}{2}\rr^\top \HH\rr+\hh^\top\rr$ and the dissipator in the form $\diss(\rho)=\sum_\mu \diss[L_\mu](\rho)$ with $L_\mu = \elll_\mu^\top \rr$, where $\elll_\mu\in\mathbb{C}^{2N}$. 
When $\rho$ is Gaussian, the Lindblad equation is translated into the following equations for mean and covariance: 
\begin{align}
    \frac{d}{dt}\rbar 
    &= \AA\rbar + \Omega \hh,\\
    \frac{d}{dt}\VV 
    &= \AA\VV + \VV\AA^\top + \DD, \label{eq:lyapunov}
\end{align}
where $\AA = \Omega \HH - \Gamma$, $\Gamma=\Omega\imag \LL$, $\DD=\Omega(\real\LL)\Omega^\top$, and $\LL = \sum_\mu \elll_\mu\elll_\mu^\dagger$~\cite{mancini2007optimal}. 

Furthermore, if a system is equipped with reference Hamiltonian $\ham_0=\omega\sum_ja_j^\dagger a_j$ and jump operators $L_{2j-1} = \sqrt{\gamma(\bar{n}+1)}a_j$ and $L_{2j} = \sqrt{\gamma\bar{n}}a_j^\dagger$ (here, $a_j = (x_j+ip_j)/\sqrt{2}$), we refer to it as a \textit{uniform $N$-mode system}. 
In this case, the dissipator reads 
\begin{align}
    \diss(\rho) 
    = \sum_j \bigg(
        \gamma(\bar{n}+1)\diss[a_j](\rho)
        + \gamma\bar{n}\diss[a_j^\dagger](\rho)
    \bigg). \label{eq:diss_N_identical}
\end{align}
For such a system, we find 
\begin{align}
    \Gamma = \frac{\gamma}{2}\id_{2N},\quad 
    \DD = \gamma\bigg(\bar{n}+\frac{1}{2}\bigg)\id_{2N}. \label{eq:gamma_and_D}
\end{align}
Thus, while $\Omega\HH$ and $\Omega\hh$ come directly from the Hamiltonian term in the dynamics, $\Gamma$ and $\DD$ represent dissipation and fluctuation residing in the dissipator. 

In this specific case, we can derive a nontrivial relation in the steady state, 
\begin{align}
    \tr(\VV^{-1}) = \frac{2N}{\bar{n}+\frac{1}{2}}. \label{eq:nontrivial_formula}
\end{align}
This is shown as follows: first, in the steady state, we have 
\begin{align}\label{eq:Lyapunov0}
    \AA\VV + \VV\AA^\top + \DD=0. 
\end{align}
By multiplying by $\VV^{-1}$ from the right and using expressions in Eq.~\eqref{eq:gamma_and_D}, we find 
\begin{align}
    \Omega \HH 
    + \VV(\Omega\HH)^\top \VV^{-1}
    - \gamma \id_{2N} 
    + \gamma \bigg(\bar{n}+\frac{1}{2}\bigg)\VV^{-1}=0. 
\end{align}
Since the inner product between a symmetric and antisymmetric matrix vanishes, $\tr(\Omega\HH)=0$ holds, and taking the trace proves Eq.~\eqref{eq:nontrivial_formula}. 

\subsection{Validity of steady-state sampling}
Here, we discuss the validity of generating a steady state of Eq.~\eqref{eq:lindblad} via Eq.~\eqref{eq:parametrization} (while it is evident that any two-dimensional vector is eligible as $\rbar$). 
We assume $N=1$. 
As discussed in Eq.~\eqref{eq:Lyapunov0}, the steady-state covariance $\VV$ satisfies 
\begin{align}
    \AA \VV + \VV\AA^\top + \DD=0 \label{eq:lyapunov_ss}
\end{align}
with $\AA=\Omega\HH-\frac{\gamma}{2}\id$ and $\DD=\gamma\big(\bar{n}+\frac{1}{2}\big)\id$. 
This equation (and the original Lindblad equation) has a stable solution if and only if $\AA$ is Hurwitz, i.e., every eigenvalue of $\AA$ has a negative real part~\cite{sontag1998mathematical}. 
Let $\mathcal{V}_{\mathrm{ss}}$ denote the set of steady-state covariance matrices obtained by varying
$\HH$ over all symmetric matrices for which $\AA=\Omega\HH-\frac{\gamma}{2}\id$ is Hurwitz, at
fixed gamma and nbar.
We also define the set of matrices parametrized as in Eq.~\eqref{eq:parametrization} by $\mathcal{V}_{\mathrm{prm}}$. 

Below we prove $\mathcal{V}_{\mathrm{ss}}=\mathcal{V}_{\mathrm{prm}}$. 
When $\VV\in\mathcal{V}_{\mathrm{ss}}$, it is given as 
\begin{align}
    \VV = \int_0^\infty e^{s\AA}\DD e^{s\AA^\top}ds. 
\end{align}
Thus, we have $\VV^\top = \VV$ and $\VV\geq0$ since $\DD^\top =\DD$ and $\DD\geq0$. 
In addition, $\VV$ must satisfy $\tr(\VV^{-1})=2/(\bar{n}+1/2)$ as already proved in Eq.~\eqref{eq:nontrivial_formula}. 
Therefore, $\VV$ is diagonalized by an orthogonal matrix, and its eigenvalues satisfy $1/v_++1/v_-=2/(\bar{n}+1/2)$. 
Since $1/v_+,1/v_-\geq 0$, they can be described by an intermediate parameter $\lambda\in[0,1)$ as 
\begin{align}
    v_{\pm} = \frac{\bar{n}+1/2}{1\mp \lambda}. 
\end{align}
By defining $s=(1/4)\ln(v_+/v_-)$, we can replace $\lambda$ with $\tanh(2s)$, which results in the parametrization in Eq.~\eqref{eq:parametrization}, showing the inclusion $\mathcal{V}_\mathrm{ss}\subseteq\mathcal{V}_{\mathrm{prm}}$. 

On the other hand, if $\VV\in\mathcal{V}_{\mathrm{prm}}$, we can show that (i) there is $\AA$ that solves Eq.~\eqref{eq:lyapunov_ss}, (ii) $\AA$ is Hurwitz, and (iii) the corresponding Hamiltonian matrix is symmetric. 
The solution is simply given by 
\begin{align}
    \AA_c = \Big(-\frac{\gamma}{2}(\bar{n}+1/2)\id + c \Omega\Big)\VV^{-1}
\end{align}
with an arbitrary parameter $c\in\mathbb{R}$. 
This is because Eq.~\eqref{eq:lyapunov_ss} turns into $\BB+\BB^\top=-\gamma(\bar{n}+1/2)\id$ when we define $\BB=\AA\VV$. 
Thus, it uses only the fact that $\VV$ is symmetric and invertible. 
Next, we can see that $\AA_c$ is Hurwitz by considering  $\tilde{\AA}_c=\VV^{-1/2}\AA\VV^{1/2}$, which has the same spectrum as $\AA_c$. 
It satisfies $\tilde{\AA}_c+\tilde{\AA}_c^\top = -\VV^{-1/2}\DD\VV^{-1/2}=-\gamma(\bar{n}+1/2)\VV^{-1}$, and for any pair of eigenvalue $\lambda$ and unit eigenvector $\zz$ of $\tilde{\AA}_c$, 
\begin{align}
    2\real(\lambda)
    = \zz^\dagger (\tilde{\AA}_c+\tilde{\AA}_c^\top)\zz
    = -\gamma(\bar{n}+1/2)\zz^\dagger\VV^{-1}\zz.
\end{align}
Because $\VV$ is positive and so is $\VV^{-1}$, this quantity is negative and $\AA_c$ is Hurwitz for any $c$. 
Finally, the corresponding Hamiltonian matrix $\HH_c$ is provided by 
\begin{align}
    \HH_c = \frac{\gamma}{2}\Omega\big[(\bar{n}+1/2)\VV^{-1}-\id\big]
    +c\VV^{-1}. 
\end{align}
Because $\tr(\VV^{-1})=2/(\bar{n}+1/2)$, we see that the symmetric matrix $(\bar{n}+1/2)\VV^{-1}-\id$ is traceless; thus, it is given as a linear combination of $\sigma_x=\begin{pmatrix}
    0&1\\1&0
\end{pmatrix}$ and $\sigma_z=\begin{pmatrix}
    1&0\\0&-1
\end{pmatrix}$. 
Now, we also have $\Omega\sigma_x=\sigma_z$ and $\Omega\sigma_z=-\sigma_x$ (note $\Omega=i\sigma_y$). 
Thus, $\HH_c$ is symmetric. 
Let us note that we used $\tr(\VV^{-1})=2/(\bar{n}+1/2)$, which follows from the parametrization. 
Therefore, $\VV$ is indeed a stable solution of Eq.~\eqref{eq:lyapunov_ss} with a physically reasonable Hamiltonian, i.e., $\VV\in\mathcal{V}_{\mathrm{ss}}$ and thus $\mathcal{V}_{\mathrm{prm}}\subseteq\mathcal{V}_{\mathrm{ss}}$. 

\subsection{Proof of Eq.~\eqref{eq:epr_formula}}
Here, we derive an extension of Equation~\eqref{eq:epr_formula} for a uniform $N$-mode system, which immediately leads to the single-mode formula. 
First, note that Eq.~\eqref{eq:epr_diss} is applicable to this case. 
Then, using $\ham_0=\omega\sum_ja_j^\dagger a_j$ and Eq.~\eqref{eq:diss_N_identical}, 
we find 
\begin{align}
    \epr
    =\gamma\beta\omega\sum_{j=1}^N\bigg(\langle a_j^\dagger a_j\rangle-\bar{n}\bigg)
    =\gamma\beta\omega\Bigg[\frac{1}{2}\langle\rr^\top\rr\rangle-N\bigg(\bar{n}+\frac{1}{2}\bigg)\Bigg]. 
\end{align}
Because $\bar{n}+1/2=\frac{1}{2}\coth\big(\frac{\beta\omega}{2}\big)$, we have  
\begin{align}
    \frac{\epr}{\gamma} 
    = \big(\langle\rr^\top\rr\rangle-N(2\bar{n}+1)\big)\arcoth(2\bar{n}+1). 
\end{align}
Now, we can erase $\bar{n}$ using Eq.~\eqref{eq:nontrivial_formula} to get 
\begin{align}
    \frac{\epr}{\gamma} 
    = \bigg(\langle\rr^\top\rr\rangle-\frac{4N^2}{\tr(\VV^{-1})}\bigg)\arcoth\bigg(\frac{4N}{\tr(\VV^{-1})}\bigg). \label{eq:em_1}
\end{align}
Because $\tr(\VV) = \langle \rr^\top\rr\rangle - \lVert\rbar\rVert^2$, 
Eq.~\eqref{eq:em_1} finally yields 
\begin{align}
    \frac{\epr}{\gamma} 
    = \bigg(\lVert\rbar\rVert^2+\frac{\tr(\VV)\tr(\VV^{-1})-4N^2}{\tr(\VV^{-1})}\bigg)\arcoth\bigg(\frac{4N}{\tr(\VV^{-1})}\bigg).
\end{align}
The positivity is verified from the Cauchy--Schwarz inequality $\tr(\VV)\tr(\VV^{-1})\geq [\tr(\id_{2N})]^2=4N^2$. 
When $N=1$, $\VV$ becomes a $2\times 2$ matrix and we can use the formula 
\begin{align}
    \tr(\VV^{-1})=\frac{\tr(\VV)}{\det(\VV)}
\end{align}
to derive Eq.~\eqref{eq:epr_formula}. 
The argument of $\arcoth$ is not smaller than one due to the identity~\eqref{eq:nontrivial_formula}. 

\clearpage

\newcommand{\lindbladeq}{2}
\newcommand{\ineqdis}{7a}
\newcommand{\ineqrot}{7b}
\newcommand{\ineqsq}{7c}
\newcommand{\ineqs}{7}
\newcommand{\fndis}{66}
\newcommand{\rotbound}{10}
\newcommand{\eprvarineq}{8}
\newcommand{\eprformula}{3}

\onecolumngrid
\begin{center}
    \textbf{\large Supplemental Material for ``\textit{Dissipation--Sensitivity Trade-Off in Dissipative Bosonic Systems}''}
    \vspace{2em}
\end{center}
\twocolumngrid

\renewcommand{\theequation}{S\arabic{equation}}
\setcounter{secnumdepth}{2}
\renewcommand{\thesection}{\Alph{section}}
\renewcommand{\thesubsection}{\thesection.\arabic{subsection}}

\section{Derivation of analytical results}
\subsection{Quantum Fisher information for single-mode Gaussian states}
We give a formula for computing the quantum Fisher information (QFI) of Gaussian states and derive the formulas 
\begin{subequations}\label{eq:app-qfi}
    \begin{align}
        \qfi_\rho(X_\dis^\phi) &= \nn_{\phi}^\top \VV^{-1}\nn_\phi, 
        \label{eq:app-qfi-dis}\\
        \qfi_\rho(X_\rot) &= \frac{\rbar^\top\VV\rbar}{\det(\VV)}
        + \frac{[\tr(\VV)]^2-4\det(\VV)}{\det(\VV)+1/4}, 
        \label{eq:app-qfi-rot}\\
        \qfi_\rho(X_\sq^\psi) &= \frac{\rbar^\top\QQ_\psi\VV\QQ_\psi\rbar}{\det(\VV)}
        + \frac{[\tr(\VV\QQ_\psi)]^2+4\det(\VV)}{\det(\VV)+1/4}. 
        \label{eq:app-qfi-sq}
    \end{align}
\end{subequations}

We begin with a pedagogical review of QFI. 
In general, we may consider the QFI $\qfi_\theta$ of a parametrized family of states $\{\rho_\theta\}$. 
It is defined as $\qfi_\theta = \tr(\rho_\theta L_\theta^2)$ with $L_\theta$ such that~\cite{braunstein1994statistical} 
\begin{align}
    \frac{d}{d\theta}\rho_\theta = \frac{1}{2}\{\rho_\theta, L_\theta\}. \label{eq:sld}
\end{align}
When we focus on $\rho_\theta = U_\theta\rho U_\theta^\dagger$ with a unitary operator $U_\theta = e^{-i\theta X}$, we find 
\begin{align}
    &\frac{d}{d\theta}\rho_\theta
    = -iU_\theta [X,\rho] U_\theta^\dagger,\\
    &\frac{1}{2}\{\rho_\theta, L_\theta\}
    = \frac{1}{2}U_\theta\{\rho,\tilde{L}_\theta\}U_\theta^\dagger, 
\end{align}
where $\tilde{L}_\theta = U_\theta^\dagger L_\theta U_\theta$ and we used $[X,U_\theta]=0$. 
Thus, if $L_\theta$ solves Eq.~\eqref{eq:sld}, $\tilde{L}_\theta$ satisfies 
\begin{align}
    \frac{1}{2}\{\rho,\tilde{L}_\theta\} 
    = -i[X,\rho]. 
\end{align}
This equation is solved by 
\begin{align}
    \tilde{L}_\theta 
    = -2i\int_0^\infty e^{-s\rho}[X,\rho]e^{-s\rho}ds, 
\end{align}
which implies that $\tilde{L}_\theta$ is indeed independent of $\theta$.
The QFI is now given as 
\begin{align}
    \qfi_\theta
    &= \tr(\rho_\theta L_\theta^2) \\
    &= \tr(\rho U_\theta^\dagger L_\theta^2 U_\theta) \\
    &= \tr(\rho \tilde{L}_\theta^2), 
\end{align}
where we used the cyclic property of the trace in the second line. 
Since $\tilde{L}_\theta$ is identical to $L$ appearing in our definition of QFI in the main text, we have shown that our definition coincides with the standard one. 

Let us consider a parametrized family of  Gaussian states $\rho_\theta$ with mean $\rbar_\theta$ and covariance $\VV_\theta$ and assume $N=1$. 
Here we again adopt $\rho_\theta = U_\theta \rho U_\theta^\dagger$ with $U_\theta = e^{-i\theta X}$ and write $\rbar_0$ and $\VV_0$ simply as $\rbar$ and $\VV$. 
Then, the QFI reads~\cite{pinel2013quantum,monras2013phase,jiang2014quantum,fadel2025quantum} 
\begin{align}
    \qfi_\rho(X) 
    = \dot{\rbar}{}^\top\VV^{-1}\dot{\rbar}
    + \frac{\tr\big[(\VV^{-1}\dot{\VV})^2\big]}{2(1+[\det(2\VV)]^{-1})}, 
\end{align}
where $\dot{\rbar}=\frac{d}{d\theta}\rbar_\theta|_{\theta =0}$ and $\dot{{\VV}}=\frac{d}{d\theta}\VV_\theta|_{\theta =0}$. 

Let us assume that the Hermitian operator $X$ is given as 
\begin{align}
    X = \frac{1}{2}\rr^\top \MM\rr 
    + \mm^\top \Omega\rr,
\end{align}
where $\MM^\top=\MM$ and $\MM$ and $\mm$ are real. 
Then, because the generator of Heisenberg time evolution reads 
\begin{align}
    i[X,\rr]
    = \Omega \MM\rr +\mm, 
\end{align}
we find 
\begin{align}
    \frac{d}{d\theta}\rbar_\theta 
    &= \Omega \MM \rbar_\theta + \mm,\\
    \frac{d}{d\theta}\VV_\theta
    &= \Omega \MM \VV_\theta + \VV_\theta \MM^\top \Omega^\top,
\end{align}
which is solved as $\rbar_\theta = \Smat_\theta \big(\rbar+\int_0^\theta d\theta'\,\Smat_{\theta'}^{-1}\mm\big)$ and $\VV_\theta = \Smat_\theta \VV \Smat_\theta^\top$ with $\Smat_\theta = e^{\theta\Omega \MM}$. 

Displacement is induced by setting $\MM=0$ and $\mm=\nn_\phi$. 
Then, we immediately find Eq.~\eqref{eq:app-qfi-dis}. 
This applies to an $N$-mode system for arbitrary $N$~\cite{pinel2013quantum,monras2013phase,jiang2014quantum}. 

Rotation and squeezing both lead to $\mm=0$. 
For a purely quadratic generator, after a lengthy calculation (found in Ref.~\cite{fadel2025quantum}), we can prove 
\begin{align}
    &\qfi_\rho\bigg(\frac{1}{2}\rr^\top\MM\rr\bigg)\notag\\
    &= \frac{\rbar^\top \MM\VV\MM\rbar}{\det(\VV)}
    + \frac{[\tr(\MM\VV)]^2-4\det(\MM)\det(\VV)}{\det(\VV)+1/4}. 
\end{align}
Applying this formula for $\MM=\id$ and $\QQ_\psi$, we obtain Eqs.~\eqref{eq:app-qfi-rot} and \eqref{eq:app-qfi-sq}. 

Here we comment on the action of $\QQ_\psi$. 
When $\MM=\QQ_\psi$, we find 
\begin{align}
    \Omega\QQ_\psi 
    &= \begin{pmatrix}
        -\cos(2\psi) & -\sin(2\psi)\\
        -\sin(2\psi) & \cos(2\psi)
    \end{pmatrix}\\
    &= -\nn_{\psi}\nn_{\psi}^\top 
    + \nn_{\psi+\pi/2}\nn_{\psi+\pi/2}^\top. 
\end{align}
Thus, it shrinks the state in direction $\psi$ and extends it in the orthogonal direction $\psi+\pi/2$, causing squeezing. 

\subsection{Detailed proofs of the dissipation--sensitivity trade-offs}
Next, we show the dissipation--sensitivity trade-offs~(\ineqs). 

First, we prove Eq.~(\ineqdis). 
Because the eigenvalues of $\VV$ are given by 
\begin{align}
    v_\pm 
    = \frac{\tr(\VV)\pm\sqrt{\tr(\VV)^2-4\det(\VV)}}{2}, 
\end{align}
the maximum value of the displacement QFI becomes 
\begin{align}
    \max_\phi\qfi_\dis(\phi) 
    = \frac{1}{v_-}
    = \frac{\tr(\VV)+\sqrt{\tr(\VV)^2-4\det(\VV)}}{2\det(\VV)}. 
\end{align}
Thus, we get 
\begin{align}
    &\qfi_\dis(\phi) - \qfi_\dis^\therm \notag\\
    &\leq 
    \frac{\tr(\VV)-2\sqrt{\det(\VV)}}{2\det(\VV)}
    +\frac{\sqrt{\tr(\VV)^2-4\det(\VV)}}{2\det(\VV)}, 
\end{align}
where we used $\qfi_{\dis}^\therm = 1/\sqrt{\det(\VV)}$. 
Although the left-hand side can be negative, as discussed in footnote~[\fndis] in the main text, we have 
\begin{align}
    \max_\phi|\qfi_\dis(\phi) - \qfi_\dis^\therm|
    = \max_\phi\qfi_\dis(\phi) - \qfi_\dis^\therm. 
\end{align}
Therefore, the left-hand side can be replaced with the absolute value. 
One further finds 
\begin{align}
    |\qfi_\dis(\phi) - \qfi_\dis^\therm|
    \leq \frac{\sqrt{\tr(\VV)^2-4\det(\VV)}}{\det(\VV)}
\end{align}
because of the inequality $x-y\leq \sqrt{x^2-y^2}$ for $x\geq y \geq 0$. This inequality is seen by 
\begin{align}
    \sqrt{x^2-y^2}
    = \sqrt{(x+y)(x-y)}
    \geq \sqrt{(x-y)^2} = x-y
\end{align}
which holds under the assumption $x\geq y\geq 0$. 
Then, Eq.~(\ineqdis) follows from the direct computation 
\begin{align}
    \frac{|\qfi_\dis(\phi) - \qfi_\dis^\therm|^2}{(\qfi_\dis^\therm)^2}
    \leq \frac{\tr(\VV)^2-4\det(\VV)}{\det(\VV)}
    \leq \frac{4\epr_\sq}{\gamma}, 
\end{align}
where we used 
\begin{align}
    \frac{\epr_\sq}{\gamma}
    &= \frac{[\tr(\VV)]^2-4\det(\VV)}{\tr(\VV)}
    \arcoth\bigg(\frac{4\det(\VV)}{\tr(\VV)}\bigg)\\
    &\geq \frac{[\tr(\VV)]^2-4\det(\VV)}{4\det(\VV)}. 
\end{align}

Since we already proved the rotation bound~(\ineqrot) in the main text, next we prove Eq.~(\ineqsq). 
For convenience, we show the QFI again 
\begin{align}
    \qfi_\sq(\psi) 
    = \frac{\rbar^\top\QQ_\psi\VV\QQ_\psi\rbar}{\det(\VV)}
    + \frac{[\tr(\VV\QQ_\psi)]^2+4\det(\VV)}{\det(\VV)+1/4} \label{eq:app-qfi_sq}
\end{align}
where 
\begin{align}
    \QQ_\psi = \begin{pmatrix}
    \sin(2\psi) & -\cos(2\psi) \\
    -\cos(2\psi) & -\sin(2\psi)\end{pmatrix}. 
\end{align}
The first term in Eq.~\eqref{eq:app-qfi_sq} is upper bounded as 
\begin{align}
    \frac{\rbar^\top\QQ_\psi\VV\QQ_\psi\rbar}{\det(\VV)}
    \leq \frac{\tr(\VV)}{\det(\VV)}\lVert\rbar\rVert^2 \label{eq:app-sqbound1}
\end{align}
because 
\begin{align}
    \rbar^\top\QQ_\psi\VV\QQ_\psi\rbar\leq \tr(\VV)\lVert\QQ_\psi\rbar\rVert^2
    = \tr(\VV)\lVert\rbar\rVert^2, 
\end{align}
where we used $\VV\leq \tr(\VV)\id$ and $\QQ_\psi^\top\QQ_\psi=\id$. 
To evaluate the second term, we use the following formula: 
\begin{align}
    [\tr(\VV\QQ_\psi)]^2 
    \leq [\tr(\VV)]^2 - 4\det(\VV). \label{eq:app-sqbound2}
\end{align}
The right-hand side is the maximum value when varying $\psi$. 
This is shown using the Pauli spectra; the space of two-dimensional symmetric real matrices is spanned by 
\begin{align}
    \sigma_x = \begin{pmatrix}
        0&1\\1&0
    \end{pmatrix},\;\;
    \sigma_z = \begin{pmatrix}
        1&0\\0&-1
    \end{pmatrix} \label{eq:app-pauli}
\end{align}
together with $\id$. 
If we write $\QQ_\psi$ and $\VV$ using this basis as 
\begin{align}
    \QQ_\psi &= \sin(2\psi)\sigma_z - \cos(2\psi)\sigma_x,\\
    \VV &= v_0\id + v_x \sigma_x+ v_z\sigma_z, 
\end{align}
we find useful relations 
\begin{align}
    &\tr(\VV) = 2v_0,\quad \det(\VV) = v_0^2-v_x^2-v_z^2, \label{eq:app-formulas}\\
    &\tr(\VV\QQ_\psi) 
    = -2v_x\cos(2\psi)+2v_z\sin(2\psi), \label{eq:app-trVQ}
\end{align}
where $2$ in the last equality comes from $\tr(\id)$. 
Equation~\eqref{eq:app-trVQ} suggests that $[\tr(\VV\QQ_\psi)]^2$ is maximized when $\tan(2\psi)=-v_z/v_x$. 
Then, we find 
\begin{align}
    [\tr(\VV\QQ_\psi)]^2
    \leq 4(v_x^2+v_z^2)
    = [\tr(\VV)]^2-4\det(\VV), 
\end{align}
where we used Eq.~\eqref{eq:app-formulas}. 

Combining Eqs.~\eqref{eq:app-sqbound1} and \eqref{eq:app-sqbound2}, the QFI is bounded by 
\begin{align}
    \qfi_{\sq}(\psi) 
    \leq \frac{\tr(\VV)}{\det(\VV)}\lVert\rbar\rVert^2
    + \frac{[\tr(\VV)]^2}{\det(\VV)+1/4}. 
\end{align}
Subtracting $\qfi_\sq^\therm = 4\det(\VV)/(\det(\VV)+1/4)$ and using $\det(\VV)+1/4\geq \det(\VV)$, we further obtain 
\begin{align}
    \qfi_{\sq}(\psi) - \qfi_\sq^\therm
    \leq \frac{\tr(\VV)}{\det(\VV)}\lVert\rbar\rVert^2
    + \frac{[\tr(\VV)]^2-4\det(\VV)}{\det(\VV)}, 
\end{align}
which is identical to the bound in Eq.~(\rotbound). Thus, Eq.~(\ineqsq) is shown in the same way. 

\subsection{Bounds on estimation precision}
Let us close this analytical section by discussing the details of Eq.~(\eprvarineq) and giving related results. 

For single parameter estimation, it is possible to saturate the Quantum Cram\'{e}r--Rao bound~\cite{paris2009quantum} 
\begin{align}
    \Var(\theta_{\mathrm{est}}) \geq \frac{1}{\qfi_\theta}. 
\end{align}
Here, the optimal unbiased estimator operator is constructed by 
\begin{align}
    E_\theta = \theta + \frac{L_\theta}{\qfi_\theta}. 
\end{align}
If we give an estimator $\theta_{\mathrm{est}}$ by constructing a POVM based on the spectral decomposition of $E_\theta$, its expectation value coincides with $\theta$, and the variance is given by 
\begin{align}
    \tr(E_\theta^2\rho_\theta) 
    - [\tr(E_\theta\rho_\theta)]^2
    = \frac{1}{\qfi_\theta}. 
\end{align}

In the setting of the main text, we can consider such an optimal estimator $\theta_{\mathrm{est}}^*$ with the least uncertainty for the thermal reference state $\rho_\therm$. 
Rotation sensing, however, does not have such an estimator because the QFI at the thermal state is zero. 
Still, the conclusion is the same. 

When $\Var(\theta_{\mathrm{est}})\leq (\qfi_\dis^\therm)^{-1}=\Var(\theta^*)$, we can apply the quantum Cram\'{e}r--Rao bound to Eq.~(\ineqdis) and obtain 
\begin{align}
    \frac{\epr_{\sq}}{\gamma}
    \geq \frac{1}{4}\bigg(\frac{[\Var(\theta_{\mathrm{est}})]^{-1}-\qfi_\dis^\therm}{\qfi_\dis^\therm}\bigg)^2. 
\end{align}
Because the optimal estimator satisfies $\qfi_\dis^\therm = [\Var(\theta^*)]^{-1}$, we finally obtain Eq.~(\eprvarineq), 
\begin{align}
    \frac{\epr_{\sq}}{\gamma}
    \geq \frac{1}{4}
    \bigg(\frac{\Var(\theta^*)}{\Var(\theta_{\mathrm{est}})}-1\bigg)^2. 
\end{align}

The quantum Cram\'{e}r--Rao bound is also applicable to Eqs.~(\ineqrot) and (\ineqsq) and leads to 
\begin{align}
    \frac{\epr}{\gamma}
    &\geq \frac{1}{4\Var(\theta_{\mathrm{est}}^\rot)},\\
    \frac{\epr}{\gamma}
    &\geq \frac{1}{4}\bigg(\frac{1}{\Var(\theta_{\mathrm{est}}^\sq)}
    -\frac{1}{\Var(\theta^{*,\sq})}\bigg), 
\end{align}
where we write $\rot$ and $\sq$ to specify the estimation task. 
Both inequalities convey the same message as Eq.~(\eprvarineq); we particularly remark the fact that, in rotation sensing, finite estimation accuracy is available only out of equilibrium ($\Var(\theta^{*,\rot})=1/\qfi_\rot^\therm=\infty$). 

\section{Details of numerical results}

\subsection{Sampling and numerical evaluation of Gaussian steady states}

To generate Fig.~3(a)--(c), we randomly sampled 8000 pairs of parameters $(\bar{n},s,\bar{x},\bar{p},\alpha,\phi,\psi)$ with $\gamma$ set to one. 
The parameters range uniformly as follows: 
\begin{equation}
    \begin{split}
        &\ln\bar{n}\in [\ln(0.02),\ln(3)],\;\; 
        s\in[0.005,1.35],\\
        &\bar{x},\bar{p}\in[-2,2],\;\;
        \alpha,\phi\in[0,\pi],\;\;
        \psi\in[0,\pi/2]
    \end{split}
\end{equation}
In Fig.~3(a), we discarded points with $\qfi_\dis(\phi)<\qfi_\dis^\therm$ because they do not represent a trade-off between sensitivity improvement and dissipation (although the bound remains valid). 

\subsection{Numerical methods for the Kerr nonlinear oscillator}
To obtain Fig.~4(a)--(c), we randomly sampled $G$, $K$, and $\bar{n}$ and solved $\liou(\rho_\infty)=-i[\ham,\rho_\infty]+\diss(\rho_\infty)=0$ with $\ham = \frac{G^*}{2}a^2+\frac{G}{2}\adsq+K\adsq a^2$ with the QuTiP package~\cite{johansson2012qutip} using the Fock cutoff. We fixed $\gamma=1$ again. 
We first generated 320 samples uniformly in the following range: 
\begin{equation}
    \begin{split}
        &\ln\bar{n}\in [\ln(0.02),\ln(1)],\;\; 
        |G|\in[0.15,1.5],\\
        &\ln|K|\in[\ln(0.05),\ln(0.6)], 
    \end{split}
\end{equation}
where $K$ can have both signs and $G$'s argument is randomly determined in $[-\pi,\pi)$. 
Then, we also considered 64 points in a less stable strong-driving/weak-nonlinearity region of $|G|\in[1.2,1.5]$ and $\ln|K|\in[\ln(0.05),\ln(0.1)]$ (if $K=0$, the steady state becomes unstable when $|G|\geq\gamma/2$~\footnote{If $K=0$, we have $\AA=\begin{pmatrix}
    \imag G - \gamma/2 & \real G \\
    \real G & - \imag G - \gamma/2
\end{pmatrix}$. Thus, the eigenvalues are $\lambda_\pm = -\gamma/2 \pm |G|$ (then we find $\lambda_+\lambda_- = (\gamma/2)^2-|G|^2$ and $\lambda_++\lambda_-=-\gamma$). Therefore, the steady state becomes unstable if $|G|>\gamma/2$. }). 
We finally tested 48 additional samples around the union of the four tightest states for each
of the three sensing tasks (12 distinct anchor states in this run). 

We begin with the Fock cutoff $N_{\mathrm{c}}=24$.
To guarantee convergence, we raise it to $32,40,48,56,64$ until the following conditions are satisfied at the cutoff $N_\mathrm{c}$: (1) the top-three-level population and energy balance error $|\langle\liou^\dagger(\ada)\rangle|$ are both below $1\times 10^{-8}$, and (2) the largest relative change from the preceding cutoff is below $5\times10^{-4}$ for $\langle\ada\rangle$, EPR, $\det(\VV)$, and all the QFIs. 

We compute the EPR and QFI based on Eq.~(\eprformula) and the formula $\qfi_\rho(X)=2\sum_{i,j|\lambda_i+\lambda_j\neq 0}\frac{(\lambda_i-\lambda_j)^2}{\lambda_i+\lambda_j}|\bra{i}X\ket{j}|^2$ for the spectral decomposition $\rho=\sum_i \lambda_i\ketbras{i}$~\cite{braunstein1994statistical}. 
For each state, we randomly chose eight angles of $\phi$ and $\psi$ and calculated the QFI. 
The plot of the displacement bound does not show points where $\qfi_\dis(\phi)<\qfi_\dis^\therm$. 
As a consequence, (a), (b), and (c) have 2962, 432, and 3456 points, respectively. 

The Example in Fig.~4 depicts the Wigner function of the steady state for $G=0.8i$, $K=0.1$, and $\bar{n}=0.05$. 
The Fock cutoff is $N_\mathrm{c}=48$. 
The Wigner function is defined by~\cite{weedbrook2012gaussian} 
\begin{align}
    &W(x,p)=\frac{1}{(2\pi)^2}\int d\bm{k}\, e^{-i\bm{k}^\top\rr}\chi(\bm{k})\notag\\ 
    &\text{with}\;\;\chi(\bm{k})=\tr[\rho e^{i\bm{k}^\top \hat{\rr}}], 
\end{align}
where we distinguished operator $\hat{\rr}$, which was written as $\rr$ up to here, from the real vector $\rr=(x,p)^\top$ by hat.


\begin{thebibliography}{84}%
\makeatletter
\providecommand \@ifxundefined [1]{%
 \@ifx{#1\undefined}
}%
\providecommand \@ifnum [1]{%
 \ifnum #1\expandafter \@firstoftwo
 \else \expandafter \@secondoftwo
 \fi
}%
\providecommand \@ifx [1]{%
 \ifx #1\expandafter \@firstoftwo
 \else \expandafter \@secondoftwo
 \fi
}%
\providecommand \natexlab [1]{#1}%
\providecommand \enquote  [1]{``#1''}%
\providecommand \bibnamefont  [1]{#1}%
\providecommand \bibfnamefont [1]{#1}%
\providecommand \citenamefont [1]{#1}%
\providecommand \href@noop [0]{\@secondoftwo}%
\providecommand \href [0]{\begingroup \@sanitize@url \@href}%
\providecommand \@href[1]{\@@startlink{#1}\@@href}%
\providecommand \@@href[1]{\endgroup#1\@@endlink}%
\providecommand \@sanitize@url [0]{\catcode `\\12\catcode `\$12\catcode `\&12\catcode `\#12\catcode `\^12\catcode `\_12\catcode `\%12\relax}%
\providecommand \@@startlink[1]{}%
\providecommand \@@endlink[0]{}%
\providecommand \url  [0]{\begingroup\@sanitize@url \@url }%
\providecommand \@url [1]{\endgroup\@href {#1}{\urlprefix }}%
\providecommand \urlprefix  [0]{URL }%
\providecommand \Eprint [0]{\href }%
\providecommand \doibase [0]{https://doi.org/}%
\providecommand \selectlanguage [0]{\@gobble}%
\providecommand \bibinfo  [0]{\@secondoftwo}%
\providecommand \bibfield  [0]{\@secondoftwo}%
\providecommand \translation [1]{[#1]}%
\providecommand \BibitemOpen [0]{}%
\providecommand \bibitemStop [0]{}%
\providecommand \bibitemNoStop [0]{.\EOS\space}%
\providecommand \EOS [0]{\spacefactor3000\relax}%
\providecommand \BibitemShut  [1]{\csname bibitem#1\endcsname}%
\let\auto@bib@innerbib\@empty
\bibitem [{\citenamefont {Oono}\ and\ \citenamefont {Paniconi}(1998)}]{oono1998steady}%
  \BibitemOpen
  \bibfield  {author} {\bibinfo {author} {\bibfnamefont {Y.}~\bibnamefont {Oono}}\ and\ \bibinfo {author} {\bibfnamefont {M.}~\bibnamefont {Paniconi}},\ }\bibfield  {title} {\bibinfo {title} {Steady state thermodynamics},\ }\href {https://academic.oup.com/ptps/article-abstract/doi/10.1143/PTPS.130.29/1842398} {\bibfield  {journal} {\bibinfo  {journal} {Prog. Theor. Phys. Suppl.}\ }\textbf {\bibinfo {volume} {130}},\ \bibinfo {pages} {29} (\bibinfo {year} {1998})}\BibitemShut {NoStop}%
\bibitem [{\citenamefont {Seifert}(2012)}]{seifert2012stochastic}%
  \BibitemOpen
  \bibfield  {author} {\bibinfo {author} {\bibfnamefont {U.}~\bibnamefont {Seifert}},\ }\bibfield  {title} {\bibinfo {title} {Stochastic thermodynamics, fluctuation theorems and molecular machines},\ }\href {https://iopscience.iop.org/article/10.1088/0034-4885/75/12/126001/meta} {\bibfield  {journal} {\bibinfo  {journal} {Rep. Prog. Phys.}\ }\textbf {\bibinfo {volume} {75}},\ \bibinfo {pages} {126001} (\bibinfo {year} {2012})}\BibitemShut {NoStop}%
\bibitem [{\citenamefont {Horowitz}\ and\ \citenamefont {Jacobs}(2015)}]{horowitz2015energy}%
  \BibitemOpen
  \bibfield  {author} {\bibinfo {author} {\bibfnamefont {J.~M.}\ \bibnamefont {Horowitz}}\ and\ \bibinfo {author} {\bibfnamefont {K.}~\bibnamefont {Jacobs}},\ }\bibfield  {title} {\bibinfo {title} {Energy cost of controlling mesoscopic quantum systems},\ }\href {https://doi.org/10.1103/PhysRevLett.115.130501} {\bibfield  {journal} {\bibinfo  {journal} {Phys. Rev. Lett.}\ }\textbf {\bibinfo {volume} {115}},\ \bibinfo {pages} {130501} (\bibinfo {year} {2015})}\BibitemShut {NoStop}%
\bibitem [{\citenamefont {Horowitz}\ \emph {et~al.}(2017)\citenamefont {Horowitz}, \citenamefont {Zhou},\ and\ \citenamefont {England}}]{horowitz2017minimum}%
  \BibitemOpen
  \bibfield  {author} {\bibinfo {author} {\bibfnamefont {J.~M.}\ \bibnamefont {Horowitz}}, \bibinfo {author} {\bibfnamefont {K.}~\bibnamefont {Zhou}},\ and\ \bibinfo {author} {\bibfnamefont {J.~L.}\ \bibnamefont {England}},\ }\bibfield  {title} {\bibinfo {title} {Minimum energetic cost to maintain a target nonequilibrium state},\ }\href {https://doi.org/10.1103/PhysRevE.95.042102} {\bibfield  {journal} {\bibinfo  {journal} {Phys. Rev. E}\ }\textbf {\bibinfo {volume} {95}},\ \bibinfo {pages} {042102} (\bibinfo {year} {2017})}\BibitemShut {NoStop}%
\bibitem [{\citenamefont {Erker}\ \emph {et~al.}(2017)\citenamefont {Erker}, \citenamefont {Mitchison}, \citenamefont {Silva}, \citenamefont {Woods}, \citenamefont {Brunner},\ and\ \citenamefont {Huber}}]{erker2017autonomous}%
  \BibitemOpen
  \bibfield  {author} {\bibinfo {author} {\bibfnamefont {P.}~\bibnamefont {Erker}}, \bibinfo {author} {\bibfnamefont {M.~T.}\ \bibnamefont {Mitchison}}, \bibinfo {author} {\bibfnamefont {R.}~\bibnamefont {Silva}}, \bibinfo {author} {\bibfnamefont {M.~P.}\ \bibnamefont {Woods}}, \bibinfo {author} {\bibfnamefont {N.}~\bibnamefont {Brunner}},\ and\ \bibinfo {author} {\bibfnamefont {M.}~\bibnamefont {Huber}},\ }\bibfield  {title} {\bibinfo {title} {Autonomous quantum clocks: Does thermodynamics limit our ability to measure time?},\ }\href {https://doi.org/10.1103/PhysRevX.7.031022} {\bibfield  {journal} {\bibinfo  {journal} {Phys. Rev. X}\ }\textbf {\bibinfo {volume} {7}},\ \bibinfo {pages} {031022} (\bibinfo {year} {2017})}\BibitemShut {NoStop}%
\bibitem [{\citenamefont {Schwarzhans}\ \emph {et~al.}(2021)\citenamefont {Schwarzhans}, \citenamefont {Lock}, \citenamefont {Erker}, \citenamefont {Friis},\ and\ \citenamefont {Huber}}]{schwarzhans2021autonomous}%
  \BibitemOpen
  \bibfield  {author} {\bibinfo {author} {\bibfnamefont {E.}~\bibnamefont {Schwarzhans}}, \bibinfo {author} {\bibfnamefont {M.~P.~E.}\ \bibnamefont {Lock}}, \bibinfo {author} {\bibfnamefont {P.}~\bibnamefont {Erker}}, \bibinfo {author} {\bibfnamefont {N.}~\bibnamefont {Friis}},\ and\ \bibinfo {author} {\bibfnamefont {M.}~\bibnamefont {Huber}},\ }\bibfield  {title} {\bibinfo {title} {Autonomous temporal probability concentration: Clockworks and the second law of thermodynamics},\ }\href {https://doi.org/10.1103/PhysRevX.11.011046} {\bibfield  {journal} {\bibinfo  {journal} {Phys. Rev. X}\ }\textbf {\bibinfo {volume} {11}},\ \bibinfo {pages} {011046} (\bibinfo {year} {2021})}\BibitemShut {NoStop}%
\bibitem [{\citenamefont {Schwarzhans}\ \emph {et~al.}(2026)\citenamefont {Schwarzhans}, \citenamefont {Apollaro}, \citenamefont {Khomchenko}, \citenamefont {Lock}, \citenamefont {Mitchison},\ and\ \citenamefont {Huber}}]{schwarzhans2026quantum}%
  \BibitemOpen
  \bibfield  {author} {\bibinfo {author} {\bibfnamefont {E.}~\bibnamefont {Schwarzhans}}, \bibinfo {author} {\bibfnamefont {T.~J.~G.}\ \bibnamefont {Apollaro}}, \bibinfo {author} {\bibfnamefont {I.}~\bibnamefont {Khomchenko}}, \bibinfo {author} {\bibfnamefont {M.~P.~E.}\ \bibnamefont {Lock}}, \bibinfo {author} {\bibfnamefont {M.~T.}\ \bibnamefont {Mitchison}},\ and\ \bibinfo {author} {\bibfnamefont {M.}~\bibnamefont {Huber}},\ }\bibfield  {title} {\bibinfo {title} {Quantum detectors as autonomous machines: Assessing the nonequilibrium thermodynamics of information acquisition},\ }\href {https://doi.org/10.1103/wm5p-tjtg} {\bibfield  {journal} {\bibinfo  {journal} {PRX Quantum}\ }\textbf {\bibinfo {volume} {7}},\ \bibinfo {pages} {033001} (\bibinfo {year} {2026})}\BibitemShut {NoStop}%
\bibitem [{\citenamefont {Sartori}\ \emph {et~al.}(2014)\citenamefont {Sartori}, \citenamefont {Granger}, \citenamefont {Lee},\ and\ \citenamefont {Horowitz}}]{sartori2014thermodynamic}%
  \BibitemOpen
  \bibfield  {author} {\bibinfo {author} {\bibfnamefont {P.}~\bibnamefont {Sartori}}, \bibinfo {author} {\bibfnamefont {L.}~\bibnamefont {Granger}}, \bibinfo {author} {\bibfnamefont {C.~F.}\ \bibnamefont {Lee}},\ and\ \bibinfo {author} {\bibfnamefont {J.~M.}\ \bibnamefont {Horowitz}},\ }\bibfield  {title} {\bibinfo {title} {Thermodynamic costs of information processing in sensory adaptation},\ }\href {https://doi.org/10.1371/journal.pcbi.1003974} {\bibfield  {journal} {\bibinfo  {journal} {PLoS Comput. Biol.}\ }\textbf {\bibinfo {volume} {10}},\ \bibinfo {pages} {e1003974} (\bibinfo {year} {2014})}\BibitemShut {NoStop}%
\bibitem [{\citenamefont {Sartori}\ and\ \citenamefont {Tu}(2015)}]{sartori2015free}%
  \BibitemOpen
  \bibfield  {author} {\bibinfo {author} {\bibfnamefont {P.}~\bibnamefont {Sartori}}\ and\ \bibinfo {author} {\bibfnamefont {Y.}~\bibnamefont {Tu}},\ }\bibfield  {title} {\bibinfo {title} {Free energy cost of reducing noise while maintaining a high sensitivity},\ }\href {https://doi.org/10.1103/PhysRevLett.115.118102} {\bibfield  {journal} {\bibinfo  {journal} {Phys. Rev. Lett.}\ }\textbf {\bibinfo {volume} {115}},\ \bibinfo {pages} {118102} (\bibinfo {year} {2015})}\BibitemShut {NoStop}%
\bibitem [{\citenamefont {Barato}\ and\ \citenamefont {Seifert}(2015)}]{barato2015thermodynamic}%
  \BibitemOpen
  \bibfield  {author} {\bibinfo {author} {\bibfnamefont {A.~C.}\ \bibnamefont {Barato}}\ and\ \bibinfo {author} {\bibfnamefont {U.}~\bibnamefont {Seifert}},\ }\bibfield  {title} {\bibinfo {title} {Thermodynamic uncertainty relation for biomolecular processes},\ }\href {https://journals.aps.org/prl/abstract/10.1103/PhysRevLett.114.158101} {\bibfield  {journal} {\bibinfo  {journal} {Phys. Rev. Lett.}\ }\textbf {\bibinfo {volume} {114}},\ \bibinfo {pages} {158101} (\bibinfo {year} {2015})}\BibitemShut {NoStop}%
\bibitem [{\citenamefont {Gingrich}\ \emph {et~al.}(2016)\citenamefont {Gingrich}, \citenamefont {Horowitz}, \citenamefont {Perunov},\ and\ \citenamefont {England}}]{gingrich2016dissipation}%
  \BibitemOpen
  \bibfield  {author} {\bibinfo {author} {\bibfnamefont {T.~R.}\ \bibnamefont {Gingrich}}, \bibinfo {author} {\bibfnamefont {J.~M.}\ \bibnamefont {Horowitz}}, \bibinfo {author} {\bibfnamefont {N.}~\bibnamefont {Perunov}},\ and\ \bibinfo {author} {\bibfnamefont {J.~L.}\ \bibnamefont {England}},\ }\bibfield  {title} {\bibinfo {title} {Dissipation bounds all steady-state current fluctuations},\ }\href {https://journals.aps.org/prl/abstract/10.1103/PhysRevLett.116.120601} {\bibfield  {journal} {\bibinfo  {journal} {Phys. Rev. Lett.}\ }\textbf {\bibinfo {volume} {116}},\ \bibinfo {pages} {120601} (\bibinfo {year} {2016})}\BibitemShut {NoStop}%
\bibitem [{\citenamefont {Horowitz}\ and\ \citenamefont {Gingrich}(2020)}]{horowitz2020thermodynamic}%
  \BibitemOpen
  \bibfield  {author} {\bibinfo {author} {\bibfnamefont {J.~M.}\ \bibnamefont {Horowitz}}\ and\ \bibinfo {author} {\bibfnamefont {T.~R.}\ \bibnamefont {Gingrich}},\ }\bibfield  {title} {\bibinfo {title} {Thermodynamic uncertainty relations constrain non-equilibrium fluctuations},\ }\href {https://www.nature.com/articles/s41567-019-0702-6} {\bibfield  {journal} {\bibinfo  {journal} {Nat. Phys.}\ }\textbf {\bibinfo {volume} {16}},\ \bibinfo {pages} {15} (\bibinfo {year} {2020})}\BibitemShut {NoStop}%
\bibitem [{\citenamefont {Dechant}(2018)}]{dechant2018multidimensional}%
  \BibitemOpen
  \bibfield  {author} {\bibinfo {author} {\bibfnamefont {A.}~\bibnamefont {Dechant}},\ }\bibfield  {title} {\bibinfo {title} {Multidimensional thermodynamic uncertainty relations},\ }\href {https://iopscience.iop.org/article/10.1088/1751-8121/aaf3ff/meta} {\bibfield  {journal} {\bibinfo  {journal} {J. Phys. A}\ }\textbf {\bibinfo {volume} {52}},\ \bibinfo {pages} {035001} (\bibinfo {year} {2018})}\BibitemShut {NoStop}%
\bibitem [{\citenamefont {Liu}\ \emph {et~al.}(2020)\citenamefont {Liu}, \citenamefont {Gong},\ and\ \citenamefont {Ueda}}]{liu2020thermodynamic}%
  \BibitemOpen
  \bibfield  {author} {\bibinfo {author} {\bibfnamefont {K.}~\bibnamefont {Liu}}, \bibinfo {author} {\bibfnamefont {Z.}~\bibnamefont {Gong}},\ and\ \bibinfo {author} {\bibfnamefont {M.}~\bibnamefont {Ueda}},\ }\bibfield  {title} {\bibinfo {title} {Thermodynamic uncertainty relation for arbitrary initial states},\ }\href {https://journals.aps.org/prl/abstract/10.1103/PhysRevLett.125.140602} {\bibfield  {journal} {\bibinfo  {journal} {Phys. Rev. Lett.}\ }\textbf {\bibinfo {volume} {125}},\ \bibinfo {pages} {140602} (\bibinfo {year} {2020})}\BibitemShut {NoStop}%
\bibitem [{\citenamefont {Van~Vu}\ and\ \citenamefont {Saito}(2022)}]{van2022thermodynamics}%
  \BibitemOpen
  \bibfield  {author} {\bibinfo {author} {\bibfnamefont {T.}~\bibnamefont {Van~Vu}}\ and\ \bibinfo {author} {\bibfnamefont {K.}~\bibnamefont {Saito}},\ }\bibfield  {title} {\bibinfo {title} {Thermodynamics of precision in markovian open quantum dynamics},\ }\href {https://doi.org/10.1103/PhysRevLett.128.140602} {\bibfield  {journal} {\bibinfo  {journal} {Phys. Rev. Lett.}\ }\textbf {\bibinfo {volume} {128}},\ \bibinfo {pages} {140602} (\bibinfo {year} {2022})}\BibitemShut {NoStop}%
\bibitem [{\citenamefont {Yoshimura}\ and\ \citenamefont {Ito}(2021)}]{yoshimura2021thermodynamic}%
  \BibitemOpen
  \bibfield  {author} {\bibinfo {author} {\bibfnamefont {K.}~\bibnamefont {Yoshimura}}\ and\ \bibinfo {author} {\bibfnamefont {S.}~\bibnamefont {Ito}},\ }\bibfield  {title} {\bibinfo {title} {Thermodynamic uncertainty relation and thermodynamic speed limit in deterministic chemical reaction networks},\ }\href {https://journals.aps.org/prl/abstract/10.1103/PhysRevLett.127.160601} {\bibfield  {journal} {\bibinfo  {journal} {Phys. Rev. Lett.}\ }\textbf {\bibinfo {volume} {127}},\ \bibinfo {pages} {160601} (\bibinfo {year} {2021})}\BibitemShut {NoStop}%
\bibitem [{\citenamefont {Brandner}\ and\ \citenamefont {Saito}(2025)}]{brandner2025thermodynamic}%
  \BibitemOpen
  \bibfield  {author} {\bibinfo {author} {\bibfnamefont {K.}~\bibnamefont {Brandner}}\ and\ \bibinfo {author} {\bibfnamefont {K.}~\bibnamefont {Saito}},\ }\bibfield  {title} {\bibinfo {title} {Thermodynamic uncertainty relations for coherent transport},\ }\href {https://doi.org/10.1103/6nww-8wcp} {\bibfield  {journal} {\bibinfo  {journal} {Phys. Rev. Lett.}\ }\textbf {\bibinfo {volume} {135}},\ \bibinfo {pages} {046302} (\bibinfo {year} {2025})}\BibitemShut {NoStop}%
\bibitem [{\citenamefont {Yoshimura}\ and\ \citenamefont {Hamazaki}(2026)}]{yoshimura2026quasiprobability}%
  \BibitemOpen
  \bibfield  {author} {\bibinfo {author} {\bibfnamefont {K.}~\bibnamefont {Yoshimura}}\ and\ \bibinfo {author} {\bibfnamefont {R.}~\bibnamefont {Hamazaki}},\ }\bibfield  {title} {\bibinfo {title} {Quasiprobability thermodynamic uncertainty relation},\ }\href {https://doi.org/10.1103/ky8n-9bcy} {\bibfield  {journal} {\bibinfo  {journal} {Phys. Rev. Lett.}\ }\textbf {\bibinfo {volume} {136}},\ \bibinfo {pages} {120406} (\bibinfo {year} {2026})}\BibitemShut {NoStop}%
\bibitem [{\citenamefont {Caves}(1981)}]{caves1981quantum}%
  \BibitemOpen
  \bibfield  {author} {\bibinfo {author} {\bibfnamefont {C.~M.}\ \bibnamefont {Caves}},\ }\bibfield  {title} {\bibinfo {title} {Quantum-mechanical noise in an interferometer},\ }\href {https://doi.org/10.1103/PhysRevD.23.1693} {\bibfield  {journal} {\bibinfo  {journal} {Phys. Rev. D}\ }\textbf {\bibinfo {volume} {23}},\ \bibinfo {pages} {1693} (\bibinfo {year} {1981})}\BibitemShut {NoStop}%
\bibitem [{\citenamefont {Schnabel}(2017)}]{schnabel2017squeezed}%
  \BibitemOpen
  \bibfield  {author} {\bibinfo {author} {\bibfnamefont {R.}~\bibnamefont {Schnabel}},\ }\bibfield  {title} {\bibinfo {title} {Squeezed states of light and their applications in laser interferometers},\ }\href {https://doi.org/10.1016/j.physrep.2017.04.001} {\bibfield  {journal} {\bibinfo  {journal} {Phys. Rep.}\ }\textbf {\bibinfo {volume} {684}},\ \bibinfo {pages} {1} (\bibinfo {year} {2017})}\BibitemShut {NoStop}%
\bibitem [{\citenamefont {Ganapathy}\ \emph {et~al.}(2023)\citenamefont {Ganapathy}, \citenamefont {Jia}, \citenamefont {Nakano} \emph {et~al.}}]{ganapathy2023broadband}%
  \BibitemOpen
  \bibfield  {author} {\bibinfo {author} {\bibfnamefont {D.}~\bibnamefont {Ganapathy}}, \bibinfo {author} {\bibfnamefont {W.}~\bibnamefont {Jia}}, \bibinfo {author} {\bibfnamefont {M.}~\bibnamefont {Nakano}}, \emph {et~al.} (\bibinfo {collaboration} {LIGO O4 Detector Collaboration}),\ }\bibfield  {title} {\bibinfo {title} {Broadband quantum enhancement of the {LIGO} detectors with frequency-dependent squeezing},\ }\href {https://doi.org/10.1103/PhysRevX.13.041021} {\bibfield  {journal} {\bibinfo  {journal} {Phys. Rev. X}\ }\textbf {\bibinfo {volume} {13}},\ \bibinfo {pages} {041021} (\bibinfo {year} {2023})}\BibitemShut {NoStop}%
\bibitem [{\citenamefont {Slusher}\ \emph {et~al.}(1985)\citenamefont {Slusher}, \citenamefont {Hollberg}, \citenamefont {Yurke}, \citenamefont {Mertz},\ and\ \citenamefont {Valley}}]{slusher1985observation}%
  \BibitemOpen
  \bibfield  {author} {\bibinfo {author} {\bibfnamefont {R.~E.}\ \bibnamefont {Slusher}}, \bibinfo {author} {\bibfnamefont {L.~W.}\ \bibnamefont {Hollberg}}, \bibinfo {author} {\bibfnamefont {B.}~\bibnamefont {Yurke}}, \bibinfo {author} {\bibfnamefont {J.~C.}\ \bibnamefont {Mertz}},\ and\ \bibinfo {author} {\bibfnamefont {J.~F.}\ \bibnamefont {Valley}},\ }\bibfield  {title} {\bibinfo {title} {Observation of squeezed states generated by four-wave mixing in an optical cavity},\ }\href {https://doi.org/10.1103/PhysRevLett.55.2409} {\bibfield  {journal} {\bibinfo  {journal} {Phys. Rev. Lett.}\ }\textbf {\bibinfo {volume} {55}},\ \bibinfo {pages} {2409} (\bibinfo {year} {1985})}\BibitemShut {NoStop}%
\bibitem [{\citenamefont {Wu}\ \emph {et~al.}(1986)\citenamefont {Wu}, \citenamefont {Kimble}, \citenamefont {Hall},\ and\ \citenamefont {Wu}}]{wu1986generation}%
  \BibitemOpen
  \bibfield  {author} {\bibinfo {author} {\bibfnamefont {L.-A.}\ \bibnamefont {Wu}}, \bibinfo {author} {\bibfnamefont {H.~J.}\ \bibnamefont {Kimble}}, \bibinfo {author} {\bibfnamefont {J.~L.}\ \bibnamefont {Hall}},\ and\ \bibinfo {author} {\bibfnamefont {H.}~\bibnamefont {Wu}},\ }\bibfield  {title} {\bibinfo {title} {Generation of squeezed states by parametric down conversion},\ }\href {https://doi.org/10.1103/PhysRevLett.57.2520} {\bibfield  {journal} {\bibinfo  {journal} {Phys. Rev. Lett.}\ }\textbf {\bibinfo {volume} {57}},\ \bibinfo {pages} {2520} (\bibinfo {year} {1986})}\BibitemShut {NoStop}%
\bibitem [{\citenamefont {Meekhof}\ \emph {et~al.}(1996)\citenamefont {Meekhof}, \citenamefont {Monroe}, \citenamefont {King}, \citenamefont {Itano},\ and\ \citenamefont {Wineland}}]{meekhof1996generation}%
  \BibitemOpen
  \bibfield  {author} {\bibinfo {author} {\bibfnamefont {D.~M.}\ \bibnamefont {Meekhof}}, \bibinfo {author} {\bibfnamefont {C.}~\bibnamefont {Monroe}}, \bibinfo {author} {\bibfnamefont {B.~E.}\ \bibnamefont {King}}, \bibinfo {author} {\bibfnamefont {W.~M.}\ \bibnamefont {Itano}},\ and\ \bibinfo {author} {\bibfnamefont {D.~J.}\ \bibnamefont {Wineland}},\ }\bibfield  {title} {\bibinfo {title} {Generation of nonclassical motional states of a trapped atom},\ }\href {https://doi.org/10.1103/PhysRevLett.76.1796} {\bibfield  {journal} {\bibinfo  {journal} {Phys. Rev. Lett.}\ }\textbf {\bibinfo {volume} {76}},\ \bibinfo {pages} {1796} (\bibinfo {year} {1996})}\BibitemShut {NoStop}%
\bibitem [{\citenamefont {Wollman}\ \emph {et~al.}(2015)\citenamefont {Wollman}, \citenamefont {Lei}, \citenamefont {Weinstein}, \citenamefont {Suh}, \citenamefont {Kronwald}, \citenamefont {Marquardt}, \citenamefont {Clerk},\ and\ \citenamefont {Schwab}}]{wollman2015quantum}%
  \BibitemOpen
  \bibfield  {author} {\bibinfo {author} {\bibfnamefont {E.~E.}\ \bibnamefont {Wollman}}, \bibinfo {author} {\bibfnamefont {C.~U.}\ \bibnamefont {Lei}}, \bibinfo {author} {\bibfnamefont {A.~J.}\ \bibnamefont {Weinstein}}, \bibinfo {author} {\bibfnamefont {J.}~\bibnamefont {Suh}}, \bibinfo {author} {\bibfnamefont {A.}~\bibnamefont {Kronwald}}, \bibinfo {author} {\bibfnamefont {F.}~\bibnamefont {Marquardt}}, \bibinfo {author} {\bibfnamefont {A.~A.}\ \bibnamefont {Clerk}},\ and\ \bibinfo {author} {\bibfnamefont {K.~C.}\ \bibnamefont {Schwab}},\ }\bibfield  {title} {\bibinfo {title} {Quantum squeezing of motion in a mechanical resonator},\ }\href {https://doi.org/10.1126/science.aac5138} {\bibfield  {journal} {\bibinfo  {journal} {Science}\ }\textbf {\bibinfo {volume} {349}},\ \bibinfo {pages} {952} (\bibinfo {year} {2015})}\BibitemShut {NoStop}%
\bibitem [{\citenamefont {Pirkkalainen}\ \emph {et~al.}(2015)\citenamefont {Pirkkalainen}, \citenamefont {Damsk{\"a}gg}, \citenamefont {Brandt}, \citenamefont {Massel},\ and\ \citenamefont {Sillanp{\"a}{\"a}}}]{pirkkalainen2015squeezing}%
  \BibitemOpen
  \bibfield  {author} {\bibinfo {author} {\bibfnamefont {J.-M.}\ \bibnamefont {Pirkkalainen}}, \bibinfo {author} {\bibfnamefont {E.}~\bibnamefont {Damsk{\"a}gg}}, \bibinfo {author} {\bibfnamefont {M.}~\bibnamefont {Brandt}}, \bibinfo {author} {\bibfnamefont {F.}~\bibnamefont {Massel}},\ and\ \bibinfo {author} {\bibfnamefont {M.~A.}\ \bibnamefont {Sillanp{\"a}{\"a}}},\ }\bibfield  {title} {\bibinfo {title} {Squeezing of quantum noise of motion in a micromechanical resonator},\ }\href {https://doi.org/10.1103/PhysRevLett.115.243601} {\bibfield  {journal} {\bibinfo  {journal} {Phys. Rev. Lett.}\ }\textbf {\bibinfo {volume} {115}},\ \bibinfo {pages} {243601} (\bibinfo {year} {2015})}\BibitemShut {NoStop}%
\bibitem [{\citenamefont {Vahlbruch}\ \emph {et~al.}(2016)\citenamefont {Vahlbruch}, \citenamefont {Mehmet}, \citenamefont {Danzmann},\ and\ \citenamefont {Schnabel}}]{vahlbruch2016detection}%
  \BibitemOpen
  \bibfield  {author} {\bibinfo {author} {\bibfnamefont {H.}~\bibnamefont {Vahlbruch}}, \bibinfo {author} {\bibfnamefont {M.}~\bibnamefont {Mehmet}}, \bibinfo {author} {\bibfnamefont {K.}~\bibnamefont {Danzmann}},\ and\ \bibinfo {author} {\bibfnamefont {R.}~\bibnamefont {Schnabel}},\ }\bibfield  {title} {\bibinfo {title} {Detection of {15 dB} squeezed states of light and their application for the absolute calibration of photoelectric quantum efficiency},\ }\href {https://doi.org/10.1103/PhysRevLett.117.110801} {\bibfield  {journal} {\bibinfo  {journal} {Phys. Rev. Lett.}\ }\textbf {\bibinfo {volume} {117}},\ \bibinfo {pages} {110801} (\bibinfo {year} {2016})}\BibitemShut {NoStop}%
\bibitem [{\citenamefont {Kamba}\ \emph {et~al.}(2025)\citenamefont {Kamba}, \citenamefont {Hara},\ and\ \citenamefont {Aikawa}}]{kamba2025quantum}%
  \BibitemOpen
  \bibfield  {author} {\bibinfo {author} {\bibfnamefont {M.}~\bibnamefont {Kamba}}, \bibinfo {author} {\bibfnamefont {N.}~\bibnamefont {Hara}},\ and\ \bibinfo {author} {\bibfnamefont {K.}~\bibnamefont {Aikawa}},\ }\bibfield  {title} {\bibinfo {title} {Quantum squeezing of a levitated nanomechanical oscillator},\ }\href {https://doi.org/10.1126/science.ady4652} {\bibfield  {journal} {\bibinfo  {journal} {Science}\ }\textbf {\bibinfo {volume} {389}},\ \bibinfo {pages} {1225} (\bibinfo {year} {2025})}\BibitemShut {NoStop}%
\bibitem [{\citenamefont {Kronwald}\ \emph {et~al.}(2013)\citenamefont {Kronwald}, \citenamefont {Marquardt},\ and\ \citenamefont {Clerk}}]{kronwald2013arbitrarily}%
  \BibitemOpen
  \bibfield  {author} {\bibinfo {author} {\bibfnamefont {A.}~\bibnamefont {Kronwald}}, \bibinfo {author} {\bibfnamefont {F.}~\bibnamefont {Marquardt}},\ and\ \bibinfo {author} {\bibfnamefont {A.~A.}\ \bibnamefont {Clerk}},\ }\bibfield  {title} {\bibinfo {title} {Arbitrarily large steady-state bosonic squeezing via dissipation},\ }\href {https://doi.org/10.1103/PhysRevA.88.063833} {\bibfield  {journal} {\bibinfo  {journal} {Phys. Rev. A}\ }\textbf {\bibinfo {volume} {88}},\ \bibinfo {pages} {063833} (\bibinfo {year} {2013})}\BibitemShut {NoStop}%
\bibitem [{\citenamefont {Kronwald}\ \emph {et~al.}(2014)\citenamefont {Kronwald}, \citenamefont {Marquardt},\ and\ \citenamefont {Clerk}}]{kronwald2014dissipative}%
  \BibitemOpen
  \bibfield  {author} {\bibinfo {author} {\bibfnamefont {A.}~\bibnamefont {Kronwald}}, \bibinfo {author} {\bibfnamefont {F.}~\bibnamefont {Marquardt}},\ and\ \bibinfo {author} {\bibfnamefont {A.~A.}\ \bibnamefont {Clerk}},\ }\bibfield  {title} {\bibinfo {title} {Dissipative optomechanical squeezing of light},\ }\href {https://doi.org/10.1088/1367-2630/16/6/063058} {\bibfield  {journal} {\bibinfo  {journal} {New J. Phys.}\ }\textbf {\bibinfo {volume} {16}},\ \bibinfo {pages} {063058} (\bibinfo {year} {2014})}\BibitemShut {NoStop}%
\bibitem [{\citenamefont {Di~Candia}\ \emph {et~al.}(2023)\citenamefont {Di~Candia}, \citenamefont {Minganti}, \citenamefont {Petrovnin}, \citenamefont {Paraoanu},\ and\ \citenamefont {Felicetti}}]{dicandia2023critical}%
  \BibitemOpen
  \bibfield  {author} {\bibinfo {author} {\bibfnamefont {R.}~\bibnamefont {Di~Candia}}, \bibinfo {author} {\bibfnamefont {F.}~\bibnamefont {Minganti}}, \bibinfo {author} {\bibfnamefont {K.~V.}\ \bibnamefont {Petrovnin}}, \bibinfo {author} {\bibfnamefont {G.~S.}\ \bibnamefont {Paraoanu}},\ and\ \bibinfo {author} {\bibfnamefont {S.}~\bibnamefont {Felicetti}},\ }\bibfield  {title} {\bibinfo {title} {Critical parametric quantum sensing},\ }\href {https://doi.org/10.1038/s41534-023-00690-z} {\bibfield  {journal} {\bibinfo  {journal} {npj Quantum Inf.}\ }\textbf {\bibinfo {volume} {9}},\ \bibinfo {pages} {23} (\bibinfo {year} {2023})}\BibitemShut {NoStop}%
\bibitem [{\citenamefont {Lei}\ \emph {et~al.}(2016)\citenamefont {Lei}, \citenamefont {Weinstein}, \citenamefont {Suh}, \citenamefont {Wollman}, \citenamefont {Kronwald}, \citenamefont {Marquardt}, \citenamefont {Clerk},\ and\ \citenamefont {Schwab}}]{lei2016quantum}%
  \BibitemOpen
  \bibfield  {author} {\bibinfo {author} {\bibfnamefont {C.~U.}\ \bibnamefont {Lei}}, \bibinfo {author} {\bibfnamefont {A.~J.}\ \bibnamefont {Weinstein}}, \bibinfo {author} {\bibfnamefont {J.}~\bibnamefont {Suh}}, \bibinfo {author} {\bibfnamefont {E.~E.}\ \bibnamefont {Wollman}}, \bibinfo {author} {\bibfnamefont {A.}~\bibnamefont {Kronwald}}, \bibinfo {author} {\bibfnamefont {F.}~\bibnamefont {Marquardt}}, \bibinfo {author} {\bibfnamefont {A.~A.}\ \bibnamefont {Clerk}},\ and\ \bibinfo {author} {\bibfnamefont {K.~C.}\ \bibnamefont {Schwab}},\ }\bibfield  {title} {\bibinfo {title} {Quantum nondemolition measurement of a quantum squeezed state beyond the 3 {dB} limit},\ }\href {https://doi.org/10.1103/PhysRevLett.117.100801} {\bibfield  {journal} {\bibinfo  {journal} {Phys. Rev. Lett.}\ }\textbf {\bibinfo {volume} {117}},\ \bibinfo {pages} {100801} (\bibinfo {year} {2016})}\BibitemShut {NoStop}%
\bibitem [{\citenamefont {Galve}\ and\ \citenamefont {Lutz}(2009)}]{galve2009nonequilibrium}%
  \BibitemOpen
  \bibfield  {author} {\bibinfo {author} {\bibfnamefont {F.}~\bibnamefont {Galve}}\ and\ \bibinfo {author} {\bibfnamefont {E.}~\bibnamefont {Lutz}},\ }\bibfield  {title} {\bibinfo {title} {Nonequilibrium thermodynamic analysis of squeezing},\ }\href {https://doi.org/10.1103/PhysRevA.79.055804} {\bibfield  {journal} {\bibinfo  {journal} {Phys. Rev. A}\ }\textbf {\bibinfo {volume} {79}},\ \bibinfo {pages} {055804} (\bibinfo {year} {2009})}\BibitemShut {NoStop}%
\bibitem [{\citenamefont {Zagoskin}\ \emph {et~al.}(2012)\citenamefont {Zagoskin}, \citenamefont {Il'ichev},\ and\ \citenamefont {Nori}}]{zagoskin2012heat}%
  \BibitemOpen
  \bibfield  {author} {\bibinfo {author} {\bibfnamefont {A.~M.}\ \bibnamefont {Zagoskin}}, \bibinfo {author} {\bibfnamefont {E.}~\bibnamefont {Il'ichev}},\ and\ \bibinfo {author} {\bibfnamefont {F.}~\bibnamefont {Nori}},\ }\bibfield  {title} {\bibinfo {title} {Heat cost of parametric generation of microwave squeezed states},\ }\href {https://doi.org/10.1103/PhysRevA.85.063811} {\bibfield  {journal} {\bibinfo  {journal} {Phys. Rev. A}\ }\textbf {\bibinfo {volume} {85}},\ \bibinfo {pages} {063811} (\bibinfo {year} {2012})}\BibitemShut {NoStop}%
\bibitem [{\citenamefont {Lipka-Bartosik}\ and\ \citenamefont {Demkowicz-Dobrza{\'{n}}ski}(2018)}]{lipkabartosik2018thermodynamic}%
  \BibitemOpen
  \bibfield  {author} {\bibinfo {author} {\bibfnamefont {P.}~\bibnamefont {Lipka-Bartosik}}\ and\ \bibinfo {author} {\bibfnamefont {R.}~\bibnamefont {Demkowicz-Dobrza{\'{n}}ski}},\ }\bibfield  {title} {\bibinfo {title} {Thermodynamic work cost of quantum estimation protocols},\ }\href {https://doi.org/10.1088/1751-8121/aae664} {\bibfield  {journal} {\bibinfo  {journal} {J. Phys. A}\ }\textbf {\bibinfo {volume} {51}},\ \bibinfo {pages} {474001} (\bibinfo {year} {2018})}\BibitemShut {NoStop}%
\bibitem [{\citenamefont {Chu}\ and\ \citenamefont {Cai}(2022{\natexlab{a}})}]{chu2022thermodynamic}%
  \BibitemOpen
  \bibfield  {author} {\bibinfo {author} {\bibfnamefont {Y.}~\bibnamefont {Chu}}\ and\ \bibinfo {author} {\bibfnamefont {J.}~\bibnamefont {Cai}},\ }\bibfield  {title} {\bibinfo {title} {Thermodynamic principle for quantum metrology},\ }\href {https://doi.org/10.1103/PhysRevLett.128.200501} {\bibfield  {journal} {\bibinfo  {journal} {Phys. Rev. Lett.}\ }\textbf {\bibinfo {volume} {128}},\ \bibinfo {pages} {200501} (\bibinfo {year} {2022}{\natexlab{a}})}\BibitemShut {NoStop}%
\bibitem [{\citenamefont {Dooley}\ \emph {et~al.}(2022)\citenamefont {Dooley}, \citenamefont {Kewming}, \citenamefont {Mitchison},\ and\ \citenamefont {Goold}}]{dooley2022comment}%
  \BibitemOpen
  \bibfield  {author} {\bibinfo {author} {\bibfnamefont {S.}~\bibnamefont {Dooley}}, \bibinfo {author} {\bibfnamefont {M.~J.}\ \bibnamefont {Kewming}}, \bibinfo {author} {\bibfnamefont {M.~T.}\ \bibnamefont {Mitchison}},\ and\ \bibinfo {author} {\bibfnamefont {J.}~\bibnamefont {Goold}},\ }\href@noop {} {\bibinfo {title} {Comment on ``{Thermodynamic Principle for Quantum Metrology}''}} (\bibinfo {year} {2022}),\ \Eprint {https://arxiv.org/abs/2205.11411} {arXiv:2205.11411} \BibitemShut {NoStop}%
\bibitem [{\citenamefont {Chu}\ and\ \citenamefont {Cai}(2022{\natexlab{b}})}]{chu2022notes}%
  \BibitemOpen
  \bibfield  {author} {\bibinfo {author} {\bibfnamefont {Y.}~\bibnamefont {Chu}}\ and\ \bibinfo {author} {\bibfnamefont {J.}~\bibnamefont {Cai}},\ }\href@noop {} {\bibinfo {title} {Notes on {Thermodynamic Principle for Quantum Metrology}}} (\bibinfo {year} {2022}{\natexlab{b}}),\ \Eprint {https://arxiv.org/abs/2208.05167} {arXiv:2208.05167} \BibitemShut {NoStop}%
\bibitem [{\citenamefont {Chu}\ and\ \citenamefont {Cai}(2022{\natexlab{c}})}]{chu2022erratum}%
  \BibitemOpen
  \bibfield  {author} {\bibinfo {author} {\bibfnamefont {Y.}~\bibnamefont {Chu}}\ and\ \bibinfo {author} {\bibfnamefont {J.}~\bibnamefont {Cai}},\ }\bibfield  {title} {\bibinfo {title} {Erratum: {Thermodynamic Principle for Quantum Metrology} [{Phys. Rev. Lett.} 128, 200501 (2022)]},\ }\href {https://doi.org/10.1103/PhysRevLett.129.109902} {\bibfield  {journal} {\bibinfo  {journal} {Phys. Rev. Lett.}\ }\textbf {\bibinfo {volume} {129}},\ \bibinfo {pages} {109902} (\bibinfo {year} {2022}{\natexlab{c}})}\BibitemShut {NoStop}%
\bibitem [{\citenamefont {Wang}\ \emph {et~al.}(2018)\citenamefont {Wang}, \citenamefont {Wu}, \citenamefont {Cui},\ and\ \citenamefont {Wang}}]{wang2018coherence}%
  \BibitemOpen
  \bibfield  {author} {\bibinfo {author} {\bibfnamefont {Z.}~\bibnamefont {Wang}}, \bibinfo {author} {\bibfnamefont {W.}~\bibnamefont {Wu}}, \bibinfo {author} {\bibfnamefont {G.}~\bibnamefont {Cui}},\ and\ \bibinfo {author} {\bibfnamefont {J.}~\bibnamefont {Wang}},\ }\bibfield  {title} {\bibinfo {title} {Coherence enhanced quantum metrology in a nonequilibrium optical molecule},\ }\href {https://doi.org/10.1088/1367-2630/aab03a} {\bibfield  {journal} {\bibinfo  {journal} {New J. Phys.}\ }\textbf {\bibinfo {volume} {20}},\ \bibinfo {pages} {033034} (\bibinfo {year} {2018})}\BibitemShut {NoStop}%
\bibitem [{\citenamefont {Wang}\ and\ \citenamefont {Wang}(2022)}]{wang2022the}%
  \BibitemOpen
  \bibfield  {author} {\bibinfo {author} {\bibfnamefont {X.}~\bibnamefont {Wang}}\ and\ \bibinfo {author} {\bibfnamefont {J.}~\bibnamefont {Wang}},\ }\bibfield  {title} {\bibinfo {title} {The effect of nonequilibrium entropy production on the quantum {Fisher} information and correlations},\ }\href {https://doi.org/10.1007/s11128-021-03277-9} {\bibfield  {journal} {\bibinfo  {journal} {Quantum Inf. Process.}\ }\textbf {\bibinfo {volume} {21}},\ \bibinfo {pages} {16} (\bibinfo {year} {2022})}\BibitemShut {NoStop}%
\bibitem [{\citenamefont {Chitgopekar}\ \emph {et~al.}(2026)\citenamefont {Chitgopekar}, \citenamefont {Koukoulekidis},\ and\ \citenamefont {Marvian}}]{chitgopekar2026thermal}%
  \BibitemOpen
  \bibfield  {author} {\bibinfo {author} {\bibfnamefont {Y.}~\bibnamefont {Chitgopekar}}, \bibinfo {author} {\bibfnamefont {N.}~\bibnamefont {Koukoulekidis}},\ and\ \bibinfo {author} {\bibfnamefont {I.}~\bibnamefont {Marvian}},\ }\href {https://arxiv.org/abs/2609.09583} {\bibinfo {title} {Thermal quantum sensing: {Fisher} information and work beyond {Gaussian} signals}} (\bibinfo {year} {2026}),\ \Eprint {https://arxiv.org/abs/2609.09583} {arXiv:2609.09583 [quant-ph]} \BibitemShut {NoStop}%
\bibitem [{\citenamefont {Giovannetti}\ \emph {et~al.}(2011)\citenamefont {Giovannetti}, \citenamefont {Lloyd},\ and\ \citenamefont {Maccone}}]{giovannetti2011advances}%
  \BibitemOpen
  \bibfield  {author} {\bibinfo {author} {\bibfnamefont {V.}~\bibnamefont {Giovannetti}}, \bibinfo {author} {\bibfnamefont {S.}~\bibnamefont {Lloyd}},\ and\ \bibinfo {author} {\bibfnamefont {L.}~\bibnamefont {Maccone}},\ }\bibfield  {title} {\bibinfo {title} {Advances in quantum metrology},\ }\href {https://doi.org/10.1038/nphoton.2011.35} {\bibfield  {journal} {\bibinfo  {journal} {Nat. Photon.}\ }\textbf {\bibinfo {volume} {5}},\ \bibinfo {pages} {222} (\bibinfo {year} {2011})}\BibitemShut {NoStop}%
\bibitem [{\citenamefont {Braunstein}\ and\ \citenamefont {Caves}(1994)}]{braunstein1994statistical}%
  \BibitemOpen
  \bibfield  {author} {\bibinfo {author} {\bibfnamefont {S.~L.}\ \bibnamefont {Braunstein}}\ and\ \bibinfo {author} {\bibfnamefont {C.~M.}\ \bibnamefont {Caves}},\ }\bibfield  {title} {\bibinfo {title} {Statistical distance and the geometry of quantum states},\ }\href {https://doi.org/10.1103/PhysRevLett.72.3439} {\bibfield  {journal} {\bibinfo  {journal} {Phys. Rev. Lett.}\ }\textbf {\bibinfo {volume} {72}},\ \bibinfo {pages} {3439} (\bibinfo {year} {1994})}\BibitemShut {NoStop}%
\bibitem [{\citenamefont {Paris}(2009)}]{paris2009quantum}%
  \BibitemOpen
  \bibfield  {author} {\bibinfo {author} {\bibfnamefont {M.~G.~A.}\ \bibnamefont {Paris}},\ }\bibfield  {title} {\bibinfo {title} {Quantum estimation for quantum technology},\ }\href {https://doi.org/10.1142/S0219749909004839} {\bibfield  {journal} {\bibinfo  {journal} {Int. J. Quantum Inf.}\ }\textbf {\bibinfo {volume} {7}},\ \bibinfo {pages} {125} (\bibinfo {year} {2009})}\BibitemShut {NoStop}%
\bibitem [{\citenamefont {Weedbrook}\ \emph {et~al.}(2012)\citenamefont {Weedbrook}, \citenamefont {Pirandola}, \citenamefont {Garc{\'{i}}a-Patr{\'{o}}n}, \citenamefont {Cerf}, \citenamefont {Ralph}, \citenamefont {Shapiro},\ and\ \citenamefont {Lloyd}}]{weedbrook2012gaussian}%
  \BibitemOpen
  \bibfield  {author} {\bibinfo {author} {\bibfnamefont {C.}~\bibnamefont {Weedbrook}}, \bibinfo {author} {\bibfnamefont {S.}~\bibnamefont {Pirandola}}, \bibinfo {author} {\bibfnamefont {R.}~\bibnamefont {Garc{\'{i}}a-Patr{\'{o}}n}}, \bibinfo {author} {\bibfnamefont {N.~J.}\ \bibnamefont {Cerf}}, \bibinfo {author} {\bibfnamefont {T.~C.}\ \bibnamefont {Ralph}}, \bibinfo {author} {\bibfnamefont {J.~H.}\ \bibnamefont {Shapiro}},\ and\ \bibinfo {author} {\bibfnamefont {S.}~\bibnamefont {Lloyd}},\ }\bibfield  {title} {\bibinfo {title} {{Gaussian} quantum information},\ }\href {https://doi.org/10.1103/RevModPhys.84.621} {\bibfield  {journal} {\bibinfo  {journal} {Rev. Mod. Phys.}\ }\textbf {\bibinfo {volume} {84}},\ \bibinfo {pages} {621} (\bibinfo {year} {2012})}\BibitemShut {NoStop}%
\bibitem [{\citenamefont {Fadel}\ \emph {et~al.}(2025)\citenamefont {Fadel}, \citenamefont {Roux},\ and\ \citenamefont {Gessner}}]{fadel2025quantum}%
  \BibitemOpen
  \bibfield  {author} {\bibinfo {author} {\bibfnamefont {M.}~\bibnamefont {Fadel}}, \bibinfo {author} {\bibfnamefont {N.}~\bibnamefont {Roux}},\ and\ \bibinfo {author} {\bibfnamefont {M.}~\bibnamefont {Gessner}},\ }\bibfield  {title} {\bibinfo {title} {Quantum metrology with a continuous-variable system},\ }\href {https://doi.org/10.1088/1361-6633/ae00d8} {\bibfield  {journal} {\bibinfo  {journal} {Rep. Prog. Phys.}\ }\textbf {\bibinfo {volume} {88}},\ \bibinfo {pages} {106001} (\bibinfo {year} {2025})}\BibitemShut {NoStop}%
\bibitem [{\citenamefont {Lindblad}(1976)}]{lindblad1976generators}%
  \BibitemOpen
  \bibfield  {author} {\bibinfo {author} {\bibfnamefont {G.}~\bibnamefont {Lindblad}},\ }\bibfield  {title} {\bibinfo {title} {On the generators of quantum dynamical semigroups},\ }\href {https://link.springer.com/article/10.1007/BF01608499} {\bibfield  {journal} {\bibinfo  {journal} {Commun. Math. Phys.}\ }\textbf {\bibinfo {volume} {48}},\ \bibinfo {pages} {119} (\bibinfo {year} {1976})}\BibitemShut {NoStop}%
\bibitem [{\citenamefont {Gorini}\ \emph {et~al.}(1976)\citenamefont {Gorini}, \citenamefont {Kossakowski},\ and\ \citenamefont {Sudarshan}}]{gorini1976completely}%
  \BibitemOpen
  \bibfield  {author} {\bibinfo {author} {\bibfnamefont {V.}~\bibnamefont {Gorini}}, \bibinfo {author} {\bibfnamefont {A.}~\bibnamefont {Kossakowski}},\ and\ \bibinfo {author} {\bibfnamefont {E.~C.~G.}\ \bibnamefont {Sudarshan}},\ }\bibfield  {title} {\bibinfo {title} {Completely positive dynamical semigroups of n-level systems},\ }\href@noop {} {\bibfield  {journal} {\bibinfo  {journal} {J. Math. Phys.}\ }\textbf {\bibinfo {volume} {17}},\ \bibinfo {pages} {821} (\bibinfo {year} {1976})}\BibitemShut {NoStop}%
\bibitem [{\citenamefont {Breuer}\ and\ \citenamefont {Petruccione}(2002)}]{breuer2002theory}%
  \BibitemOpen
  \bibfield  {author} {\bibinfo {author} {\bibfnamefont {H.-P.}\ \bibnamefont {Breuer}}\ and\ \bibinfo {author} {\bibfnamefont {F.}~\bibnamefont {Petruccione}},\ }\href@noop {} {\emph {\bibinfo {title} {The theory of open quantum systems}}}\ (\bibinfo  {publisher} {Oxford University Press, New York},\ \bibinfo {year} {2002})\BibitemShut {NoStop}%
\bibitem [{\citenamefont {Milburn}\ and\ \citenamefont {Walls}(1981)}]{milburn1981production}%
  \BibitemOpen
  \bibfield  {author} {\bibinfo {author} {\bibfnamefont {G.}~\bibnamefont {Milburn}}\ and\ \bibinfo {author} {\bibfnamefont {D.}~\bibnamefont {Walls}},\ }\bibfield  {title} {\bibinfo {title} {Production of squeezed states in a degenerate parametric amplifier},\ }\href {https://doi.org/10.1016/0030-4018(81)90232-7} {\bibfield  {journal} {\bibinfo  {journal} {Opt. Commun.}\ }\textbf {\bibinfo {volume} {39}},\ \bibinfo {pages} {401} (\bibinfo {year} {1981})}\BibitemShut {NoStop}%
\bibitem [{\citenamefont {Collett}\ and\ \citenamefont {Gardiner}(1984)}]{collett1984squeezing}%
  \BibitemOpen
  \bibfield  {author} {\bibinfo {author} {\bibfnamefont {M.}~\bibnamefont {Collett}}\ and\ \bibinfo {author} {\bibfnamefont {C.}~\bibnamefont {Gardiner}},\ }\bibfield  {title} {\bibinfo {title} {Squeezing of intracavity and traveling-wave light fields produced in parametric amplification},\ }\href {https://doi.org/10.1103/PhysRevA.30.1386} {\bibfield  {journal} {\bibinfo  {journal} {Phys. Rev. A}\ }\textbf {\bibinfo {volume} {30}},\ \bibinfo {pages} {1386} (\bibinfo {year} {1984})}\BibitemShut {NoStop}%
\bibitem [{\citenamefont {Alicki}(1979)}]{alicki1979quantum}%
  \BibitemOpen
  \bibfield  {author} {\bibinfo {author} {\bibfnamefont {R.}~\bibnamefont {Alicki}},\ }\bibfield  {title} {\bibinfo {title} {The quantum open system as a model of the heat engine},\ }\href {https://doi.org/10.1088/0305-4470/12/5/007} {\bibfield  {journal} {\bibinfo  {journal} {J. Phys. A: Math. Gen.}\ }\textbf {\bibinfo {volume} {12}},\ \bibinfo {pages} {L103} (\bibinfo {year} {1979})}\BibitemShut {NoStop}%
\bibitem [{\citenamefont {Spohn}(1978)}]{spohn1978entropy}%
  \BibitemOpen
  \bibfield  {author} {\bibinfo {author} {\bibfnamefont {H.}~\bibnamefont {Spohn}},\ }\bibfield  {title} {\bibinfo {title} {Entropy production for quantum dynamical semigroups},\ }\href {https://doi.org/10.1063/1.523789} {\bibfield  {journal} {\bibinfo  {journal} {J. Math. Phys.}\ }\textbf {\bibinfo {volume} {19}},\ \bibinfo {pages} {1227} (\bibinfo {year} {1978})}\BibitemShut {NoStop}%
\bibitem [{\citenamefont {Landi}\ and\ \citenamefont {Paternostro}(2021)}]{landi2021irreversible}%
  \BibitemOpen
  \bibfield  {author} {\bibinfo {author} {\bibfnamefont {G.~T.}\ \bibnamefont {Landi}}\ and\ \bibinfo {author} {\bibfnamefont {M.}~\bibnamefont {Paternostro}},\ }\bibfield  {title} {\bibinfo {title} {Irreversible entropy production: From classical to quantum},\ }\href {https://journals.aps.org/rmp/abstract/10.1103/RevModPhys.93.035008} {\bibfield  {journal} {\bibinfo  {journal} {Rev. Mod. Phys.}\ }\textbf {\bibinfo {volume} {93}},\ \bibinfo {pages} {035008} (\bibinfo {year} {2021})}\BibitemShut {NoStop}%
\bibitem [{\citenamefont {Santos}\ \emph {et~al.}(2017)\citenamefont {Santos}, \citenamefont {Landi},\ and\ \citenamefont {Paternostro}}]{santos2017wigner}%
  \BibitemOpen
  \bibfield  {author} {\bibinfo {author} {\bibfnamefont {J.~P.}\ \bibnamefont {Santos}}, \bibinfo {author} {\bibfnamefont {G.~T.}\ \bibnamefont {Landi}},\ and\ \bibinfo {author} {\bibfnamefont {M.}~\bibnamefont {Paternostro}},\ }\bibfield  {title} {\bibinfo {title} {{Wigner} entropy production rate},\ }\href {https://doi.org/10.1103/PhysRevLett.118.220601} {\bibfield  {journal} {\bibinfo  {journal} {Phys. Rev. Lett.}\ }\textbf {\bibinfo {volume} {118}},\ \bibinfo {pages} {220601} (\bibinfo {year} {2017})}\BibitemShut {NoStop}%
\bibitem [{\citenamefont {Mancini}\ and\ \citenamefont {Wiseman}(2007)}]{mancini2007optimal}%
  \BibitemOpen
  \bibfield  {author} {\bibinfo {author} {\bibfnamefont {S.}~\bibnamefont {Mancini}}\ and\ \bibinfo {author} {\bibfnamefont {H.~M.}\ \bibnamefont {Wiseman}},\ }\bibfield  {title} {\bibinfo {title} {Optimal control of entanglement via quantum feedback},\ }\href {https://doi.org/10.1103/PhysRevA.75.012330} {\bibfield  {journal} {\bibinfo  {journal} {Phys. Rev. A}\ }\textbf {\bibinfo {volume} {75}},\ \bibinfo {pages} {012330} (\bibinfo {year} {2007})}\BibitemShut {NoStop}%
\bibitem [{\citenamefont {Rudnicki}\ and\ \citenamefont {Gneiting}(2018)}]{rudnicki2018stabilizable}%
  \BibitemOpen
  \bibfield  {author} {\bibinfo {author} {\bibfnamefont {{\L}.}~\bibnamefont {Rudnicki}}\ and\ \bibinfo {author} {\bibfnamefont {C.}~\bibnamefont {Gneiting}},\ }\bibfield  {title} {\bibinfo {title} {Stabilizable {Gaussian} states},\ }\href {https://doi.org/10.1103/PhysRevA.98.032120} {\bibfield  {journal} {\bibinfo  {journal} {Phys. Rev. A}\ }\textbf {\bibinfo {volume} {98}},\ \bibinfo {pages} {032120} (\bibinfo {year} {2018})}\BibitemShut {NoStop}%
\bibitem [{\citenamefont {Manzano}\ \emph {et~al.}(2016)\citenamefont {Manzano}, \citenamefont {Galve}, \citenamefont {Zambrini},\ and\ \citenamefont {Parrondo}}]{manzano2016entropy}%
  \BibitemOpen
  \bibfield  {author} {\bibinfo {author} {\bibfnamefont {G.}~\bibnamefont {Manzano}}, \bibinfo {author} {\bibfnamefont {F.}~\bibnamefont {Galve}}, \bibinfo {author} {\bibfnamefont {R.}~\bibnamefont {Zambrini}},\ and\ \bibinfo {author} {\bibfnamefont {J.~M.}\ \bibnamefont {Parrondo}},\ }\bibfield  {title} {\bibinfo {title} {Entropy production and thermodynamic power of the squeezed thermal reservoir},\ }\href {https://doi.org/10.1103/PhysRevE.93.052120} {\bibfield  {journal} {\bibinfo  {journal} {Phys. Rev. E}\ }\textbf {\bibinfo {volume} {93}},\ \bibinfo {pages} {052120} (\bibinfo {year} {2016})}\BibitemShut {NoStop}%
\bibitem [{\citenamefont {Singh}\ and\ \citenamefont {M{\"u}stecapl{\i}o{\u{g}}lu}(2020)}]{singh2020performance}%
  \BibitemOpen
  \bibfield  {author} {\bibinfo {author} {\bibfnamefont {V.}~\bibnamefont {Singh}}\ and\ \bibinfo {author} {\bibfnamefont {{\"O}.~E.}\ \bibnamefont {M{\"u}stecapl{\i}o{\u{g}}lu}},\ }\bibfield  {title} {\bibinfo {title} {Performance bounds of nonadiabatic quantum harmonic otto engine and refrigerator under a squeezed thermal reservoir},\ }\href {https://doi.org/10.1103/PhysRevE.102.062123} {\bibfield  {journal} {\bibinfo  {journal} {Phys. Rev. E}\ }\textbf {\bibinfo {volume} {102}},\ \bibinfo {pages} {062123} (\bibinfo {year} {2020})}\BibitemShut {NoStop}%
\bibitem [{smA()}]{smArxiv}%
  \BibitemOpen
  \href@noop {} {}\bibinfo {note} {Supplemental Material after the End Matter, in which we discuss the details of analytical results and numerical calculations.}\BibitemShut {Stop}%
\bibitem [{\citenamefont {Pinel}\ \emph {et~al.}(2013)\citenamefont {Pinel}, \citenamefont {Jian}, \citenamefont {Treps}, \citenamefont {Fabre},\ and\ \citenamefont {Braun}}]{pinel2013quantum}%
  \BibitemOpen
  \bibfield  {author} {\bibinfo {author} {\bibfnamefont {O.}~\bibnamefont {Pinel}}, \bibinfo {author} {\bibfnamefont {P.}~\bibnamefont {Jian}}, \bibinfo {author} {\bibfnamefont {N.}~\bibnamefont {Treps}}, \bibinfo {author} {\bibfnamefont {C.}~\bibnamefont {Fabre}},\ and\ \bibinfo {author} {\bibfnamefont {D.}~\bibnamefont {Braun}},\ }\bibfield  {title} {\bibinfo {title} {Quantum parameter estimation using general single-mode {Gaussian} states},\ }\href {https://doi.org/10.1103/PhysRevA.88.040102} {\bibfield  {journal} {\bibinfo  {journal} {Phys. Rev. A}\ }\textbf {\bibinfo {volume} {88}},\ \bibinfo {pages} {040102} (\bibinfo {year} {2013})}\BibitemShut {NoStop}%
\bibitem [{\citenamefont {Monras}(2013)}]{monras2013phase}%
  \BibitemOpen
  \bibfield  {author} {\bibinfo {author} {\bibfnamefont {A.}~\bibnamefont {Monras}},\ }\bibfield  {title} {\bibinfo {title} {Phase space formalism for quantum estimation of gaussian states},\ }\href {https://arxiv.org/abs/1303.3682} {\bibfield  {journal} {\bibinfo  {journal} {arXiv preprint}\ } (\bibinfo {year} {2013})},\ \Eprint {https://arxiv.org/abs/1303.3682} {arXiv:1303.3682 [quant-ph]} \BibitemShut {NoStop}%
\bibitem [{\citenamefont {Jiang}(2014)}]{jiang2014quantum}%
  \BibitemOpen
  \bibfield  {author} {\bibinfo {author} {\bibfnamefont {Z.}~\bibnamefont {Jiang}},\ }\bibfield  {title} {\bibinfo {title} {Quantum fisher information for states in exponential form},\ }\href {https://doi.org/10.1103/PhysRevA.89.032128} {\bibfield  {journal} {\bibinfo  {journal} {Phys. Rev. A}\ }\textbf {\bibinfo {volume} {89}},\ \bibinfo {pages} {032128} (\bibinfo {year} {2014})}\BibitemShut {NoStop}%
\bibitem [{\citenamefont {Serafini}(2023)}]{serafini2023quantum}%
  \BibitemOpen
  \bibfield  {author} {\bibinfo {author} {\bibfnamefont {A.}~\bibnamefont {Serafini}},\ }\href@noop {} {\emph {\bibinfo {title} {Quantum continuous variables: a primer of theoretical methods}}}\ (\bibinfo  {publisher} {CRC press},\ \bibinfo {year} {2023})\BibitemShut {NoStop}%
\bibitem [{Note1()}]{Note1}%
  \BibitemOpen
  \bibinfo {note} {With $v_+\geq v_-$ being the eigenvalues of $\protect \mathbb {V}$, the QFI falls in the interval $[1/v_+,1/v_-]$, while $\protect \mathcal {F}_\protect \mathrm {dis}^{\protect \mathrm {th}}=1/{\protect \sqrt {v_+v_-}}$. Thus, we get $\min _\phi \protect \mathcal {F}_\protect \mathrm {dis}(\phi )\leq \protect \mathcal {F}_\protect \mathrm {dis}^{\protect \mathrm {th}}\leq \max _\phi \protect \mathcal {F}_\protect \mathrm {dis}(\phi )$. Due to the inequality between the arithmetic mean and the geometric mean, we also have $[\max _\phi \protect \mathcal {F}_\protect \mathrm {dis}(\phi )+\min _\phi \protect \mathcal {F}_\protect \mathrm {dis}(\phi )]/2\geq \protect \mathcal {F}_\protect \mathrm {dis}^{\protect \mathrm {th}}$, which means the lower bound in Eq.~\protect \eqref {eq:ineq_dis} is maximized when $\phi =\arg \max _{\phi '}\protect \mathcal {F}_\protect \mathrm {dis}(\phi ')$ (seen by drawing a parabola).}\BibitemShut {Stop}%
\bibitem [{\citenamefont {Kwon}\ \emph {et~al.}(2019)\citenamefont {Kwon}, \citenamefont {Tan}, \citenamefont {Volkoff},\ and\ \citenamefont {Jeong}}]{kwon2019nonclassicality}%
  \BibitemOpen
  \bibfield  {author} {\bibinfo {author} {\bibfnamefont {H.}~\bibnamefont {Kwon}}, \bibinfo {author} {\bibfnamefont {K.~C.}\ \bibnamefont {Tan}}, \bibinfo {author} {\bibfnamefont {T.}~\bibnamefont {Volkoff}},\ and\ \bibinfo {author} {\bibfnamefont {H.}~\bibnamefont {Jeong}},\ }\bibfield  {title} {\bibinfo {title} {Nonclassicality as a quantifiable resource for quantum metrology},\ }\href {https://doi.org/10.1103/PhysRevLett.122.040503} {\bibfield  {journal} {\bibinfo  {journal} {Phys. Rev. Lett.}\ }\textbf {\bibinfo {volume} {122}},\ \bibinfo {pages} {040503} (\bibinfo {year} {2019})}\BibitemShut {NoStop}%
\bibitem [{\citenamefont {Lloyd}\ and\ \citenamefont {Braunstein}(1999)}]{lloyd1999quantum}%
  \BibitemOpen
  \bibfield  {author} {\bibinfo {author} {\bibfnamefont {S.}~\bibnamefont {Lloyd}}\ and\ \bibinfo {author} {\bibfnamefont {S.~L.}\ \bibnamefont {Braunstein}},\ }\bibfield  {title} {\bibinfo {title} {Quantum computation over continuous variables},\ }\href {https://doi.org/10.1103/PhysRevLett.82.1784} {\bibfield  {journal} {\bibinfo  {journal} {Phys. Rev. Lett.}\ }\textbf {\bibinfo {volume} {82}},\ \bibinfo {pages} {1784} (\bibinfo {year} {1999})}\BibitemShut {NoStop}%
\bibitem [{\citenamefont {Walschaers}(2021)}]{walschaers2021non}%
  \BibitemOpen
  \bibfield  {author} {\bibinfo {author} {\bibfnamefont {M.}~\bibnamefont {Walschaers}},\ }\bibfield  {title} {\bibinfo {title} {Non-gaussian quantum states and where to find them},\ }\href {https://doi.org/10.1103/PRXQuantum.2.030204} {\bibfield  {journal} {\bibinfo  {journal} {PRX Quantum}\ }\textbf {\bibinfo {volume} {2}},\ \bibinfo {pages} {030204} (\bibinfo {year} {2021})}\BibitemShut {NoStop}%
\bibitem [{\citenamefont {Drummond}\ and\ \citenamefont {Walls}(1980)}]{drummond1980quantum}%
  \BibitemOpen
  \bibfield  {author} {\bibinfo {author} {\bibfnamefont {P.}~\bibnamefont {Drummond}}\ and\ \bibinfo {author} {\bibfnamefont {D.}~\bibnamefont {Walls}},\ }\bibfield  {title} {\bibinfo {title} {Quantum theory of optical bistability. i. nonlinear polarisability model},\ }\href {https://doi.org/10.1088/0305-4470/13/2/034} {\bibfield  {journal} {\bibinfo  {journal} {J. Phys. A: Math. Gen.}\ }\textbf {\bibinfo {volume} {13}},\ \bibinfo {pages} {725} (\bibinfo {year} {1980})}\BibitemShut {NoStop}%
\bibitem [{\citenamefont {Kirchmair}\ \emph {et~al.}(2013)\citenamefont {Kirchmair}, \citenamefont {Vlastakis}, \citenamefont {Leghtas}, \citenamefont {Nigg}, \citenamefont {Paik}, \citenamefont {Ginossar}, \citenamefont {Mirrahimi}, \citenamefont {Frunzio}, \citenamefont {Girvin},\ and\ \citenamefont {Schoelkopf}}]{kirchmair2013observation}%
  \BibitemOpen
  \bibfield  {author} {\bibinfo {author} {\bibfnamefont {G.}~\bibnamefont {Kirchmair}}, \bibinfo {author} {\bibfnamefont {B.}~\bibnamefont {Vlastakis}}, \bibinfo {author} {\bibfnamefont {Z.}~\bibnamefont {Leghtas}}, \bibinfo {author} {\bibfnamefont {S.~E.}\ \bibnamefont {Nigg}}, \bibinfo {author} {\bibfnamefont {H.}~\bibnamefont {Paik}}, \bibinfo {author} {\bibfnamefont {E.}~\bibnamefont {Ginossar}}, \bibinfo {author} {\bibfnamefont {M.}~\bibnamefont {Mirrahimi}}, \bibinfo {author} {\bibfnamefont {L.}~\bibnamefont {Frunzio}}, \bibinfo {author} {\bibfnamefont {S.~M.}\ \bibnamefont {Girvin}},\ and\ \bibinfo {author} {\bibfnamefont {R.~J.}\ \bibnamefont {Schoelkopf}},\ }\bibfield  {title} {\bibinfo {title} {Observation of quantum state collapse and revival due to the single-photon kerr effect},\ }\href {https://doi.org/10.1038/nature11902} {\bibfield  {journal} {\bibinfo  {journal} {Nature}\ }\textbf {\bibinfo {volume} {495}},\ \bibinfo {pages} {205} (\bibinfo {year} {2013})}\BibitemShut {NoStop}%
\bibitem [{\citenamefont {Shiraishi}\ \emph {et~al.}(2018)\citenamefont {Shiraishi}, \citenamefont {Funo},\ and\ \citenamefont {Saito}}]{shiraishi2018speed}%
  \BibitemOpen
  \bibfield  {author} {\bibinfo {author} {\bibfnamefont {N.}~\bibnamefont {Shiraishi}}, \bibinfo {author} {\bibfnamefont {K.}~\bibnamefont {Funo}},\ and\ \bibinfo {author} {\bibfnamefont {K.}~\bibnamefont {Saito}},\ }\bibfield  {title} {\bibinfo {title} {Speed limit for classical stochastic processes},\ }\href {https://journals.aps.org/prl/abstract/10.1103/PhysRevLett.121.070601} {\bibfield  {journal} {\bibinfo  {journal} {Phys. Rev. Lett.}\ }\textbf {\bibinfo {volume} {121}},\ \bibinfo {pages} {070601} (\bibinfo {year} {2018})}\BibitemShut {NoStop}%
\bibitem [{\citenamefont {Funo}\ \emph {et~al.}(2019)\citenamefont {Funo}, \citenamefont {Shiraishi},\ and\ \citenamefont {Saito}}]{funo2019speed}%
  \BibitemOpen
  \bibfield  {author} {\bibinfo {author} {\bibfnamefont {K.}~\bibnamefont {Funo}}, \bibinfo {author} {\bibfnamefont {N.}~\bibnamefont {Shiraishi}},\ and\ \bibinfo {author} {\bibfnamefont {K.}~\bibnamefont {Saito}},\ }\bibfield  {title} {\bibinfo {title} {Speed limit for open quantum systems},\ }\href {https://doi.org/10.1088/1367-2630/aaf9f5} {\bibfield  {journal} {\bibinfo  {journal} {New J. Phys.}\ }\textbf {\bibinfo {volume} {21}},\ \bibinfo {pages} {013006} (\bibinfo {year} {2019})}\BibitemShut {NoStop}%
\bibitem [{\citenamefont {Ito}\ and\ \citenamefont {Dechant}(2020)}]{ito2020stochastic}%
  \BibitemOpen
  \bibfield  {author} {\bibinfo {author} {\bibfnamefont {S.}~\bibnamefont {Ito}}\ and\ \bibinfo {author} {\bibfnamefont {A.}~\bibnamefont {Dechant}},\ }\bibfield  {title} {\bibinfo {title} {Stochastic time evolution, information geometry, and the cram{\'e}r-rao bound},\ }\href {https://journals.aps.org/prx/abstract/10.1103/PhysRevX.10.021056} {\bibfield  {journal} {\bibinfo  {journal} {Phys. Rev. X}\ }\textbf {\bibinfo {volume} {10}},\ \bibinfo {pages} {021056} (\bibinfo {year} {2020})}\BibitemShut {NoStop}%
\bibitem [{\citenamefont {Hamazaki}(2022)}]{hamazaki2022speed}%
  \BibitemOpen
  \bibfield  {author} {\bibinfo {author} {\bibfnamefont {R.}~\bibnamefont {Hamazaki}},\ }\bibfield  {title} {\bibinfo {title} {Speed limits for macroscopic transitions},\ }\href {https://doi.org/10.1103/PRXQuantum.3.020319} {\bibfield  {journal} {\bibinfo  {journal} {PRX Quantum}\ }\textbf {\bibinfo {volume} {3}},\ \bibinfo {pages} {020319} (\bibinfo {year} {2022})}\BibitemShut {NoStop}%
\bibitem [{\citenamefont {Tajima}\ and\ \citenamefont {Funo}(2021)}]{tajima2021superconducting}%
  \BibitemOpen
  \bibfield  {author} {\bibinfo {author} {\bibfnamefont {H.}~\bibnamefont {Tajima}}\ and\ \bibinfo {author} {\bibfnamefont {K.}~\bibnamefont {Funo}},\ }\bibfield  {title} {\bibinfo {title} {Superconducting-like heat current: Effective cancellation of current-dissipation trade-off by quantum coherence},\ }\href {https://journals.aps.org/prl/abstract/10.1103/PhysRevLett.127.190604} {\bibfield  {journal} {\bibinfo  {journal} {Phys. Rev. Lett.}\ }\textbf {\bibinfo {volume} {127}},\ \bibinfo {pages} {190604} (\bibinfo {year} {2021})}\BibitemShut {NoStop}%
\bibitem [{\citenamefont {Funo}\ and\ \citenamefont {Tajima}(2025)}]{funo2025symmetry}%
  \BibitemOpen
  \bibfield  {author} {\bibinfo {author} {\bibfnamefont {K.}~\bibnamefont {Funo}}\ and\ \bibinfo {author} {\bibfnamefont {H.}~\bibnamefont {Tajima}},\ }\bibfield  {title} {\bibinfo {title} {Symmetry induced enhancement in finite-time thermodynamic trade-off relations},\ }\href {https://doi.org/10.1103/PhysRevLett.134.080401} {\bibfield  {journal} {\bibinfo  {journal} {Phys. Rev. Lett.}\ }\textbf {\bibinfo {volume} {134}},\ \bibinfo {pages} {080401} (\bibinfo {year} {2025})}\BibitemShut {NoStop}%
\bibitem [{\citenamefont {Kamimura}\ \emph {et~al.}(2022)\citenamefont {Kamimura}, \citenamefont {Hakoshima}, \citenamefont {Matsuzaki}, \citenamefont {Yoshida},\ and\ \citenamefont {Tokura}}]{kamimura2022quantum}%
  \BibitemOpen
  \bibfield  {author} {\bibinfo {author} {\bibfnamefont {S.}~\bibnamefont {Kamimura}}, \bibinfo {author} {\bibfnamefont {H.}~\bibnamefont {Hakoshima}}, \bibinfo {author} {\bibfnamefont {Y.}~\bibnamefont {Matsuzaki}}, \bibinfo {author} {\bibfnamefont {K.}~\bibnamefont {Yoshida}},\ and\ \bibinfo {author} {\bibfnamefont {Y.}~\bibnamefont {Tokura}},\ }\bibfield  {title} {\bibinfo {title} {Quantum-enhanced heat engine based on superabsorption},\ }\href {https://doi.org/10.1103/PhysRevLett.128.180602} {\bibfield  {journal} {\bibinfo  {journal} {Phys. Rev. Lett.}\ }\textbf {\bibinfo {volume} {128}},\ \bibinfo {pages} {180602} (\bibinfo {year} {2022})}\BibitemShut {NoStop}%
\bibitem [{\citenamefont {Kamimura}\ \emph {et~al.}(2023)\citenamefont {Kamimura}, \citenamefont {Yoshida}, \citenamefont {Tokura},\ and\ \citenamefont {Matsuzaki}}]{kamimura2023universal}%
  \BibitemOpen
  \bibfield  {author} {\bibinfo {author} {\bibfnamefont {S.}~\bibnamefont {Kamimura}}, \bibinfo {author} {\bibfnamefont {K.}~\bibnamefont {Yoshida}}, \bibinfo {author} {\bibfnamefont {Y.}~\bibnamefont {Tokura}},\ and\ \bibinfo {author} {\bibfnamefont {Y.}~\bibnamefont {Matsuzaki}},\ }\bibfield  {title} {\bibinfo {title} {Universal scaling bounds on a quantum heat current},\ }\href {https://doi.org/10.1103/PhysRevLett.131.090401} {\bibfield  {journal} {\bibinfo  {journal} {Phys. Rev. Lett.}\ }\textbf {\bibinfo {volume} {131}},\ \bibinfo {pages} {090401} (\bibinfo {year} {2023})}\BibitemShut {NoStop}%
\bibitem [{\citenamefont {Johansson}\ \emph {et~al.}(2012)\citenamefont {Johansson}, \citenamefont {Nation},\ and\ \citenamefont {Nori}}]{johansson2012qutip}%
  \BibitemOpen
  \bibfield  {author} {\bibinfo {author} {\bibfnamefont {J.~R.}\ \bibnamefont {Johansson}}, \bibinfo {author} {\bibfnamefont {P.~D.}\ \bibnamefont {Nation}},\ and\ \bibinfo {author} {\bibfnamefont {F.}~\bibnamefont {Nori}},\ }\bibfield  {title} {\bibinfo {title} {Qutip: An open-source python framework for the dynamics of open quantum systems},\ }\href {https://www.sciencedirect.com/science/article/pii/S0010465512000835?casa_token=7PgMV9h4Qm4AAAAA:OY4kghYuzAmsPP2NqbGz37iMPwDnJspHX7VNgIy2UZcMBH2pUb0W9Puuf_jVTIP-NplkfXKvCQ} {\bibfield  {journal} {\bibinfo  {journal} {Comput. Phys. Commun.}\ }\textbf {\bibinfo {volume} {183}},\ \bibinfo {pages} {1760} (\bibinfo {year} {2012})}\BibitemShut {NoStop}%
\bibitem [{dat()}]{data}%
  \BibitemOpen
  \href@noop {} {}\bibinfo {note} {\url{https://github.com/ykoheiuwu/DissipationSensitivityTradeoff}}\BibitemShut {NoStop}%
\bibitem [{\citenamefont {Medina}\ \emph {et~al.}(2025)\citenamefont {Medina}, \citenamefont {Culhane}, \citenamefont {Binder}, \citenamefont {Landi},\ and\ \citenamefont {Goold}}]{medina2025anomalous}%
  \BibitemOpen
  \bibfield  {author} {\bibinfo {author} {\bibfnamefont {I.}~\bibnamefont {Medina}}, \bibinfo {author} {\bibfnamefont {O.}~\bibnamefont {Culhane}}, \bibinfo {author} {\bibfnamefont {F.~C.}\ \bibnamefont {Binder}}, \bibinfo {author} {\bibfnamefont {G.~T.}\ \bibnamefont {Landi}},\ and\ \bibinfo {author} {\bibfnamefont {J.}~\bibnamefont {Goold}},\ }\bibfield  {title} {\bibinfo {title} {Anomalous discharging of quantum batteries: The ergotropic mpemba effect},\ }\href {https://doi.org/10.1103/PhysRevLett.134.220402} {\bibfield  {journal} {\bibinfo  {journal} {Phys. Rev. Lett.}\ }\textbf {\bibinfo {volume} {134}},\ \bibinfo {pages} {220402} (\bibinfo {year} {2025})}\BibitemShut {NoStop}%
\bibitem [{\citenamefont {Sontag}\ \emph {et~al.}(1998)\citenamefont {Sontag} \emph {et~al.}}]{sontag1998mathematical}%
  \BibitemOpen
  \bibfield  {author} {\bibinfo {author} {\bibfnamefont {E.~D.}\ \bibnamefont {Sontag}} \emph {et~al.},\ }\href@noop {} {\emph {\bibinfo {title} {Mathematical control theory}}},\ Vol.~\bibinfo {volume} {2}\ (\bibinfo  {publisher} {Springer-Verlag New York},\ \bibinfo {year} {1998})\BibitemShut {NoStop}%
\bibitem [{Note2()}]{Note2}%
  \BibitemOpen
  \bibinfo {note} {If $K=0$, we have $\protect \mathbb {A}=\begin {pmatrix} \mathop {\protect \mathrm {Im}}G - \gamma /2 & \mathop {\protect \mathrm {Re}}G \\ \mathop {\protect \mathrm {Re}}G & - \mathop {\protect \mathrm {Im}}G - \gamma /2 \end {pmatrix}$. Thus, the eigenvalues are $\lambda _\pm = -\gamma /2 \pm |G|$ (then we find $\lambda _+\lambda _- = (\gamma /2)^2-|G|^2$ and $\lambda _++\lambda _-=-\gamma $). Therefore, the steady state becomes unstable if $|G|>\gamma /2$.}\BibitemShut {Stop}%
\end{thebibliography}
\end{document}